\documentclass{aa}  

\usepackage{graphicx}
\usepackage{txfonts}
\usepackage{hyperref}
\usepackage{color}
\usepackage{upgreek}
\usepackage{tikz}

\begin{document}

   \title{Radio observations of the merging galaxy cluster Abell 520}


   \author{D. N. Hoang\inst{1}, T. W. Shimwell\inst{2,1}, R. J. van Weeren\inst{1}, G. Brunetti\inst{3}, H. J. A. R\"{o}ttgering\inst{1}, F. Andrade-Santos\inst{4}, \mbox{A. Botteon\inst{3,5}}, M. Br\"uggen\inst{6}, R. Cassano\inst{3}, A. Drabent\inst{7}, F. de Gasperin\inst{6}, M. Hoeft\inst{7}, H. T. Intema\inst{1}, D. A. Rafferty\inst{6}, \mbox{A. Shweta}\inst{8}, and A. Stroe\inst{9}    
          }

   \institute{Leiden Observatory, Leiden University, PO Box 9513, NL-2300 RA Leiden, the Netherlands \\
              \email{hoang@strw.leidenuniv.nl}
         \and
             Netherlands Institute for Radio Astronomy (ASTRON), P.O. Box 2, 7990 AA Dwingeloo, The Netherlands %
                \and
                        INAF-Istituto di Radioastronomia, via P. Gobetti 101, 40129, Bologna, Italy 
                \and
                        Harvard-Smithsonian for Astrophysics, 60 Garden Street, Cambridge, MA 02138, USA                 
                \and
                        Dipartimento di Fisica e Astronomia, Universit\`{a} di Bologna, via P. Gobetti 93/2, I-40129 Bologna, Italy 
                \and
                        Hamburger Sternwarte, University of Hamburg, Gojenbergsweg 112, 21029 Hamburg, Germany 
                \and
                        Th\"uringer Landessternwarte, Sternwarte 5, D-07778 Tautenburg, Germany.
                \and
                        Indian Institute of Science Education and Research (IISER) Pune, India
                \and
                        European Southern Observatory, Karl-Schwarzschild-Str. 2, D-85748 Garching, Germany 
             }
   \titlerunning{Radio observations of Abell 520}
   \authorrunning{D. N. Hoang et. al.}
   \date{Received..., 2018; ..., 2018}
        
 
  \abstract
   {Extended synchrotron radio sources are often observed in merging galaxy clusters. Studies of the extended emission help us to understand the mechanisms in which the radio emitting particles gain their relativistic energies.}
        {We examine the possible acceleration mechanisms of the relativistic particles that are responsible for the extended radio emission in the merging galaxy cluster Abell 520.}
   {We performed new 145 MHz observations with the LOw Frequency ARay  (LOFAR) and combined these with archival Giant Metrewave Radio Telescope (GMRT) 323 MHz and Very Large Array (VLA) 1.5 GHz data to study the morphological and spectral properties of extended cluster emission. The observational properties are discussed in the framework of particle acceleration models associated with cluster merger turbulence and shocks.}
   {In Abell 520, we confirm the presence of extended ($760\times950\,\text{kpc}^2$) synchrotron radio emission that has been classified as a radio halo. The comparison between the radio and \mbox{X-ray} brightness suggests that the halo might originate in a cocoon rather than from the central \mbox{X-ray} bright regions of the cluster. The halo spectrum is roughly uniform on the scale of $66\,\text{kpc}$. There is a hint of spectral steepening from the SW edge towards the cluster centre. Assuming diffusive shock acceleration (DSA), the radio data are suggestive of a shock Mach number of $\mathcal{M}_\text{SW}=2.6_{-0.2}^{+0.3}$ that is consistent with the \mbox{X-ray} derived estimates. This is in agreement with the scenario in which relativistic electrons in the SW radio edge gain their energies at the shock front via  acceleration of either thermal or fossil electrons. We do not detect extended radio emission ahead of the SW shock that is predicted if the emission is the result of adiabatic compression. An \mbox{X-ray} surface brightness discontinuity is detected towards the NE region that may be a  counter shock of Mach number $\mathcal{M}_\text{NE}^\text{X}=1.52\pm0.05$. This is lower than the value predicted from the radio emission which, assuming DSA, is consistent with $\mathcal{M}_\text{NE}=2.1\pm0.2$.}
  {Our observations indicate that the radio emission in the SW of Abell 520 is likely effected by the prominent \mbox{X-ray} detected shock in which radio emitting particles are \mbox{(re-)accelerated} through the Fermi-I mechanism. The NE \mbox{X-ray} discontinuity that is approximately collocated with an edge in the radio emission hints at the presence of a counter shock.}


   \maketitle
%

\section{Introduction}
\label{sec:intro}

Non-thermal components, i.e. relativistic particles and magnetic fields, in the intra-cluster medium (ICM) are important tracers of the formation and evolution of large-scale structures. The origin of these components and the role they play in the physical processes in the ICM during cluster mergers are still being investigated. There is evidence that extended synchrotron emission, namely haloes and relics, can be generated during the mergers of sub-clusters and groups during which a part of the gravitational energy goes into particle acceleration and  amplification of large-scale magnetic fields \citep[for reviews see][]{Bruggen2012,Luigina2012,Brunetti2014}. 

Radio haloes are megaparsec-scale, faint synchrotron sources that are approximately co-spatial with the thermal emission from the ICM. At moderate observing resolutions, these haloes are measured to be unpolarised down to a few percent at $\sim\text{GHz}$ frequencies. The mechanism of particle acceleration is powered by turbulence that is introduced  during a merging event  \citep[e.g.][]{Brunetti2001,Petrosian2001a,Fujita2003,Cassano2005,Brunetti2007a,Brunetti2016,Pinzke2017}. Hadronic cosmic ray (CR) proton-proton collisions in the ICM may also contribute to the observed radiation as they produce secondary relativistic electrons  \citep[e.g.][]{Dennison1980a,Blasi1999,Dolag2000,Miniati2001,Pfrommer2004a,Pfrommer2008,Keshet2010,EnBlin2011}. However, current limits from the Fermi-LAT severely challenge a scenario of pure hadronic models for radio haloes  \citep[e.g.][]{Jeltema2011,Brunetti2012,Zandanel2014a,Ackermann2010,Ackermann2016b}, thereby leaving open the possibility of hadronic models in which secondary particles are \mbox{re-accelerated} by turbulence \cite{Brunetti2011,Brunetti2017,Pinzke2017}.

Radio relics are faint, elongated synchrotron sources in the peripheral regions of galaxy clusters. These relics can have projected sizes up to $\sim\text{megaparsecs}$ and are often measured to have a high degree of polarisation of up to 70 percent. Relics are thought to form from merger or accretion shocks that accelerate particles to relativistic energies via, for example the \mbox{Fermi-I} diffusive shock acceleration \citep[DSA; e.g.][]{Ensslin1998,Roettiger1999a,Ensslin2001,Pfrommer2008}. However, the merger shocks with Mach number of $\mathcal{M}\lesssim5$ might be insufficient to generate the observed brightness and spectra of relics in a number of clusters \citep[e.g.][]{Stroe2013a,weeren2013,VanWeeren2016b,VanWeeren2017,Bonafede2014,Akamatsu2015,shimwell2015,Vazza2015,Botteon2016a,Hoang2017a,Hoang2018a}. To overcome this problem, a pre-existing population of fossil electrons that is \mbox{re-accelerated} at shock is required to be present prior to the shock passage   \citep[e.g.][]{Markevitch2005,Kang2011a,Kang2012}.

Studies of the processes occurring at cluster shock fronts are best carried out using systems for which accurate observational constraints can be derived. The best observations require relatively strong shocks ($\mathcal{M}\lesssim3$), which are usually found in the peripheral regions of galaxy clusters where the electron density is very low ($\lesssim10^{-4}\,\text{cm}^{-3}$). However, in these regions the low gas density can make it challenging to observe temperature and surface brightness (SB) discontinuities with the current generation of \mbox{X-ray} telescopes \citep{Markevitch2007,Botteon2018a}. Additionally, to best study the shocks it is preferable to minimise the mixing of different populations of electrons along the line of sight and large spherical shocks are preferred, since for these shocks the de-projection is more accurate. As merger induced shocks in clusters are rare because the systems must be caught within a limited stage of the overall merging event, there are only a handful of known strong shocks ($\mathcal{M}_{X}\sim2-3$) that meet these conditions, namely those in Abell 520 \citep{ Markevitch2005}, Abell 2146 \cite{Russell2011,Russell2012}, Abell 665 \citep{Dasadia2016}, Abell 115 \citep[e.g.][]{Botteon2016a}, and El Gordo \citep[][]{Botteon2016b}. 

\section{Galaxy cluster Abell 520}
\label{sec:a520}

The Train Wreck Cluster Abell 520 (hereafter A520; $z=0.201$) is a highly disturbed merging galaxy cluster with a merger axis $\sim60^\circ$ to the plane of the sky \citep[e.g.][]{Proust2000,Mahdavi2007,Girardi2008,Markevitch2005,Govoni2001c,Jee2012,Vacca2014}. The total mass for A520 is estimated to be $M=7.8\times10^{14}\,M_{\odot}$ in \cite{Planck2015}. Previous studies at radio wavelengths have revealed $\sim\text{Mpc}$-scale, faint ($\sim\upmu\text{Jy\,arcsec}^{-2}$ at $1.4\,\text{GHz}$) emission associated with the ICM, which has been classified as a radio halo \citep[e.g.][]{Giovannini1999,Govoni2001c,Vacca2014}. The spectral energy distribution was found to be patchy with a mean value of $-1.25$\footnote{The convention $S\propto\nu^\alpha$ is used in this paper.} and a dispersion of $0.22$  according to observations at a resolution of $39\arcsec$ \citep[][]{Vacca2014}. In the SW region, the level of radio emission rapidly drops at the location of a prominent $\mathcal{M}_{X}=2.1_{-0.3}^{+0.4}$ bow shock, where the shock strength was derived from SB and temperature jumps from \textit{Chandra} \mbox{X-ray} data \citep{Markevitch2005,Wang2016,Wang2018a}. In addition there may be another shock towards the NE of the cluster \citep{Wang2016} but the radio emission in this region has not been carefully examined. Edges of radio haloes, such as that in the SW of A520, at the locations of shocks have been observed in several other clusters \citep[][]{Markevitch2010,brown2011b,Macario2011a,Shimwell2016a,VanWeeren2016b}. However, a few of these situations, including in Abell 520, are puzzling since the expected properties of a radio relic, such as spectral transversal steepening gradients, have not been detected  \citep{Vacca2014}. 

In this paper, our main aim is to examine the low-frequency emission from the cluster and to use these measurements to better constrain the spectral energy distribution of the diffuse synchrotron emission. This allows us to search for a spectral index structure associated with the shock in the SW. In addition we can also examine the structure of the radio emission in the NE at the location of the possible shock front identified by \cite{Wang2016}. To achieve this aim, we observed the cluster with the LOw Frequency ARray (LOFAR; \citealt{VanHaarlem2013}) using high band antennas (HBA, $120-168\,\text{MHz}$) and combined these data with existing Giant Metrewave Radio Telescope (GMRT) $306-339\,\text{MHz}$ and Very Large Array (VLA) $1-2\,\text{GHz}$ data. We also make use of archival \textit{Chandra} \mbox{X-ray} data.

In this study, we assume $H_{0}=70\,\text{km\,s}^{-1}\,\text{Mpc}^{-1}$, $\Omega_{M}=0.3$ and $\Omega_{\Lambda}=0.7$. In this cosmology, an angular distance of $1\,\text{arcmin}$ corresponds to a physical scale of $198.78$ kpc at the cluster redshift of $z=0.201$.

\section{Observations and data reduction}
\label{sec:obs_red}

\subsection{LOFAR 145 MHz}
\label{sec:red_lofar}

The LOFAR 145 MHz observations of A520 were performed for a total of 7.3 hours divided equally between 17- 25 April 2017 (project: LC7\_025). The calibrator \mbox{3C 147} was observed for 10 minutes. A summary of the observations is given in \mbox{Table \ref{tab:obs}}.
\begin{table*}
        \centering
        \caption{Radio observations of A520 }   
        \begin{tabular}{lccc}
                \hline\hline
                Telescope                       &          LOFAR 145 MHz  &             GMRT 323 MHz             &              VLA 1.5 GHz \\ \hline
                Project                         &         LC7\_25         &           27\_070            &      AF349, AC706, AC776      \\
                Observation IDs                 &    L584441, L589773     &          7394, 8007          &          \textemdash          \\
                Configuration                   &       \textemdash       &         \textemdash          &             C, D              \\
                Calibrator                      &         \mbox{3C 147}          &            \mbox{3C 147}            & \mbox{3C 48}, \mbox{3C 147}, \mbox{3C 138}, 3C 286 \\
                Observation dates               &  Apr. 17 and 25, 2017   & Oct. 31, 2014; Aug. 21, 2015 &   1998 Dec. 8; 1999 Mar. 19   \\
                                                &                         &                              &   2004 Aug. 30; 2005 Aug 30   \\
                Total on-source time (hr)       &           7.3           &             17.2             &             15.1              \\
                Correlations                    &       full Stokes       &            RR, LL            &          full Stokes          \\
                Bandwidth (MHz)                 &           48            &              33              &              150              \\
                Channel width (MHz)             &         0.0122          &            0.1302            &              50               \\
                Time resolution (s)             &            1            &              16              &              10               \\
                Number of (used) stations       &           62            &              28              &            $24-27$            \\ \hline\hline
        \end{tabular}
        \label{tab:obs}
\end{table*}

The calibration of the LOFAR data was carried out using the facet calibration scheme to correct for the direction-independent and direction-dependent effects, which are implemented in the $\texttt{PreFactor}$\footnote{\url{https://github.com/lofar-astron/prefactor}} and $\texttt{Factor}$\footnote{\url{https://github.com/lofar-astron/factor}} pipelines. The data reduction procedure is described in detail in \cite{VanWeeren2016a}, \cite{Williams2016a}, and \cite{deGasperin2018a}. In particular, the data were flagged for radio interference frequency (RFI) with $\texttt{Aoflagger}$ \citep{Offringa2012c},  were removed the contamination of the bright sources in the distant side lobes (i.e. Cassiopeia A, Taurus A), and were corrected for the initial phase offsets between the  XX and YY polarisations. The clock offsets of different stations were also removed. The flux scale of the target data was calibrated according to the \citet{Scaife2012} flux scale using the primary calibrator \mbox{3C 147}. In $\texttt{Factor}$, the data were corrected for  direction-dependent distortions that are mainly caused by ionospheric effects and errors in the beam model. After this pipeline processing, the final calibrated data from the different observations were combined for imaging (see details in Sec. \ref{sec:red_im_spx}).

\subsection{GMRT 323 MHz}
\label{sec:red_gmrt}

A520 was observed with the GMRT 323 MHz on 31 October 2014 and 21 August 2015 for a total of 17.2 hours (project: 27\_070; PI: A. Shweta and R. Athreya). The calibrator \mbox{3C 147} was observed before and after the target. The observation details are summarised in Table \ref{tab:obs}. 

The calibration of the GMRT data was carried out in $\texttt{SPAM}$ \citep[Source Peeling and Atmospheric Modelling;][]{Intema2009a}. In this procedure the absolute flux scale was calibrated using  \mbox{3C 147} and a source model consistent with the \cite{Scaife2012} flux density scale. The data were flagged for RFI, and the gain and bandpass were calibrated. The direction-dependent calibration was  performed with multiple self-calibration loops to correct for the ionospheric phase delay towards the direction of the target. The final calibrated data were used to make continuum images of A520 (see Sec. \ref{sec:red_im_spx} for details).

\subsection{VLA 1.5 GHz}
\label{sec:red_vla}
We combined multiple archival L-band data sets centred on A520. These data were observed in C and D configurations. The data were originally presented in \cite{Govoni2001c} (project: AF349) and \cite{Vacca2014} (project: AC776 and AC706). Details of the observations are summarised in Table \ref{tab:obs}.

The VLA data are separately calibrated in $\texttt{CASA}$ using the flux calibrator \mbox{3C 48} for project AF349, \mbox{3C 147} for project AC776, \mbox{3C 147}, \mbox{3C 138}, and 3C 286 for project AC706. The phase calibrator is $0459+024$ for project AF349 and $0503+020$ for projects AC706 and AC776. The amplitude is calibrated according to the \cite{Perley2013} flux scale, which has an uncertainty of a few percent for these calibrators. After the initial calibration, the data are self-calibrated with  phase-only calibration steps before phase-amplitude calibration steps. The calibrated data from all observations are combined in the \textit{(u,v)}-plane and used to make continuum images of A520  (see Sec. \ref{sec:red_im_spx}).

\subsection{Continuum imaging and spectrum mapping}
\label{sec:red_im_spx}

To map the diffuse emission from A520, the LOFAR, GMRT, and VLA calibrated data sets were deconvolved with the $\texttt{MS-MFS}$ (multi-scale and multi-frequency synthesis) $\texttt{CLEAN}$ algorithm in $\texttt{CASA}$ (\citealt{Mullin2007,Cornwell2008,Rau2011}). The $\texttt{MS-MFS}$ option was used to properly model frequency-dependent emission and more accurately deconvolve extended objects. A wide-field algorithm \citep[$\texttt{W-projection}$;,][]{Cornwell2005,Cornwell2008} was also used to account for the baseline non-coplanarity over the sky. The diffuse emission at different spatial scales was enhanced using \cite{briggs1995} weighting schemes with multiple $\texttt{robust}$ values (see Table \ref{tab:image_para}). The primary beam correction for LOFAR was performed by dividing the image by the square root of the $\texttt{.avgpb}$ map  generated by $\texttt{AWimager}$  \citep{Tasse2013}. Whereas, the GMRT images were divided by a primary beam approximated by \footnote{GMRT User's manual}
\begin{equation}
A(x) = 1-\frac{3.397}{10^3}x^2+\frac{47.192}{10^7}x^4-\frac{30.931}{10^{10}}x^6+\frac{7.803}{10^{13}}x^8,
\end{equation}
where $x=f\times\theta$ with $f=0.323$ GHz and angular distance $\theta$ from the pointing centre in arcminutes. The VLA images were primary beam corrected using the built-in $\texttt{pbcor}$ option of  the $\texttt{CASA}$ $\texttt{CLEAN}$ task.

\begin{table*}
        \centering
        \caption{Image properties}
        \begin{tabular}{lcccccc}
                \hline\hline
                Telescope & \textit{uv}-range & $\texttt{robust}^a$ &      $\texttt{outertaper}$      &            resolution            &           $\sigma$            &                  Fig.                  \\
                          &   ($k\lambda$)    &                     &            ($\arcsec$)             &        (${\arcsec}^2$, p.a.)        & ($\upmu\text{Jy\,beam}^{-1}$) &                                        \\ \hline
                LOFAR     & $\geqslant0.100$  &       0.0        &               15                & $31.3\times20.4$ ($47.7^\circ$)  &              460              &          \ref{fig:a520_mres}           \\
                          &  $0.114-16.220$   &       0.0        &                5                &         $20\times 20^b$          &              450              &           \ref{fig:spx}$^c$            \\
                GMRT      & $\geqslant0.100$  &       0.0       & 15 & $18.3\times15.5$ ($60.8^\circ$)                                   &              100              &          \ref{fig:a520_mres}           \\
                          &  $0.114-16.220$   &       0.0        &               16                &          $20\times20^b$          &              130              &           \ref{fig:spx}$^c$            \\
                VLA       &  $0.114-16.220$   &       0.0        &                5                & $19.5\times18.7$ ($-43.1^\circ$) &              26               & \ref{fig:a520_mres}, \ref{fig:spx}$^c$  \\ \hline \hline
        \end{tabular} \\ 
        Notes: $^a$: Briggs weighting of \textit{uv} data; $^b$: smoothed; $^c$: spectral index map
        \label{tab:image_para}
\end{table*}

Spectral index maps of A520 were made with the LOFAR, GMRT, and VLA Stokes I images that were made using similar imaging parameters (i.e. \textit{uv}-range, $\texttt{MS-MFS}$, $\texttt{W-projection}$ options, $\texttt{Briggs}$' $\texttt{robust}$ weighting; see Table \ref{tab:image_para}). Additionally, we used an $\texttt{outertaper}$ to adjust the weightings of long baselines to obtain a spatial resolution of $\sim20\,\arcsec$ for each of the maps. The LOFAR, GMRT, and LOFAR Stokes I images were then smoothed with 2D Gaussians to a common resolution of exactly $20\,\arcsec$.  These images were also corrected for any astrometric misalignments between the images. To do this, we fit 2D Gaussians to compact sources in the images. The central positions of the Gaussians were considered as approximate locations of the sources. The misalignments between the images was taken to be the mean of the offsets between the locations of the compact sources. The images were then regridded to be identical in size. To calculate the spectral index map, only the $\geq3\sigma$ pixels that were detected in at least two images were used. These were fit with a power-law function of the form $S\propto\nu^\alpha$. The spectral index errors were calculated taking into account the image noise and a flux scale uncertainty of $15\%$ for LOFAR, $10\%$ for GMRT, and $5\%$ for VLA.

We note that to ideally recover the same spatial scales at each frequency the \textit{uv}-coverage of each data set should be matched. Generally, this matching is approximated by imaging the data sets with a uniform weighting scheme and with a common \textit{uv}-coverage in each data set, but this can result in a significantly increased noise level. Therefore, to maintain a sufficient S/N detection of A520 we have instead imaged the data with equal \textit{uv}-range and a combination of a robust parameter and taper to ensure that the images produced are of approximately equal resolution.

\subsection{\textit{Chandra}}
\label{sec:red_chandra}

Archival \textit{Chandra} data in the energy band of $0.5-2.0$ keV were fully calibrated and are published as part of an \mbox{X-ray} sample study in \cite{Andrade-Santos2017} from which we acquired the calibrated data. The observation IDs of the archival data used are 528, 4215, 7703, 9424, 9425, 9426, and 9430, resulting in a total observing duration of 528 ks. 

\section{Results}
\label{sec:res}

\begin{figure*}
        \centering
        \includegraphics[width=0.32\textwidth]{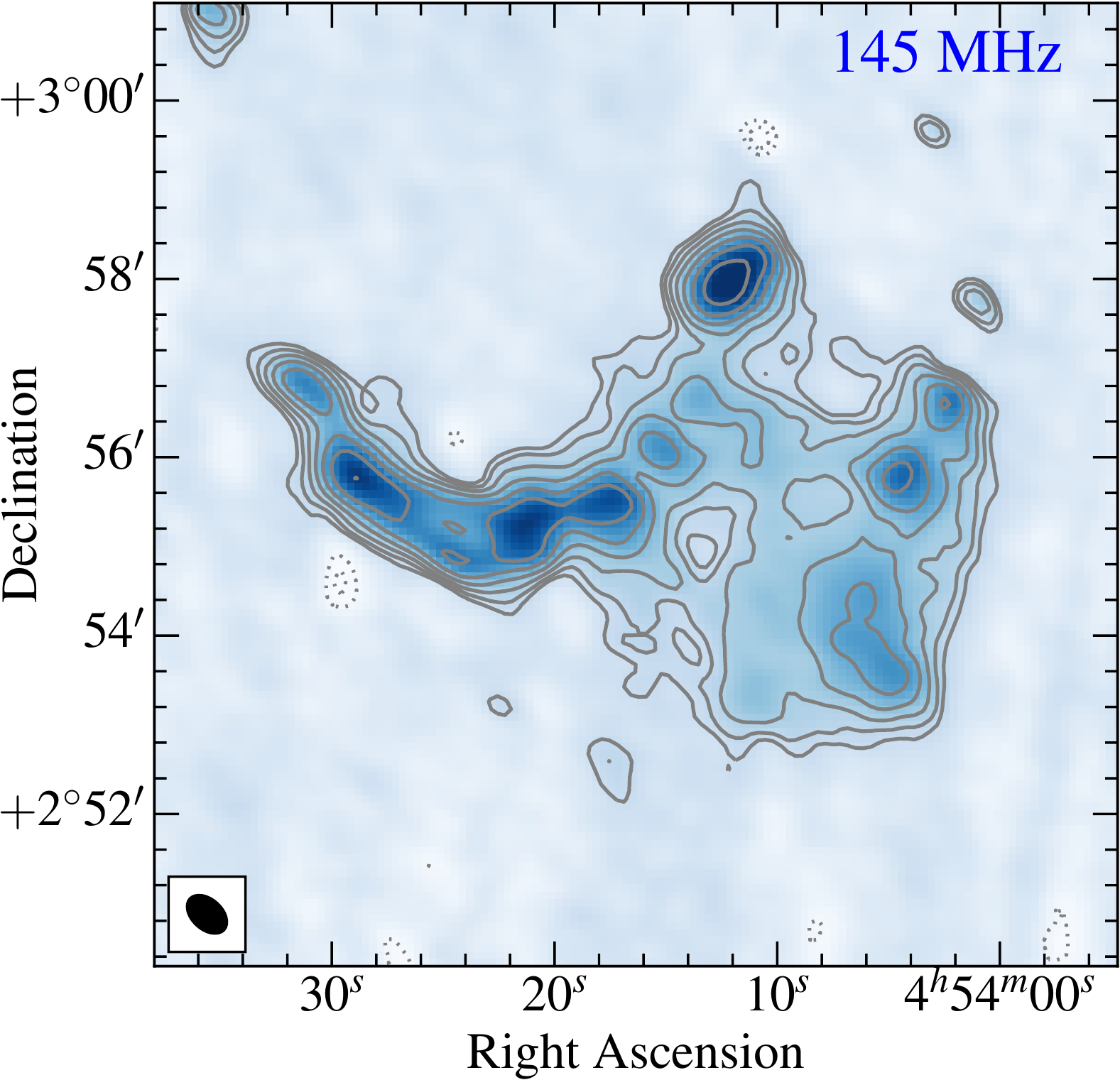}\hfill
        \includegraphics[width=0.32\textwidth]{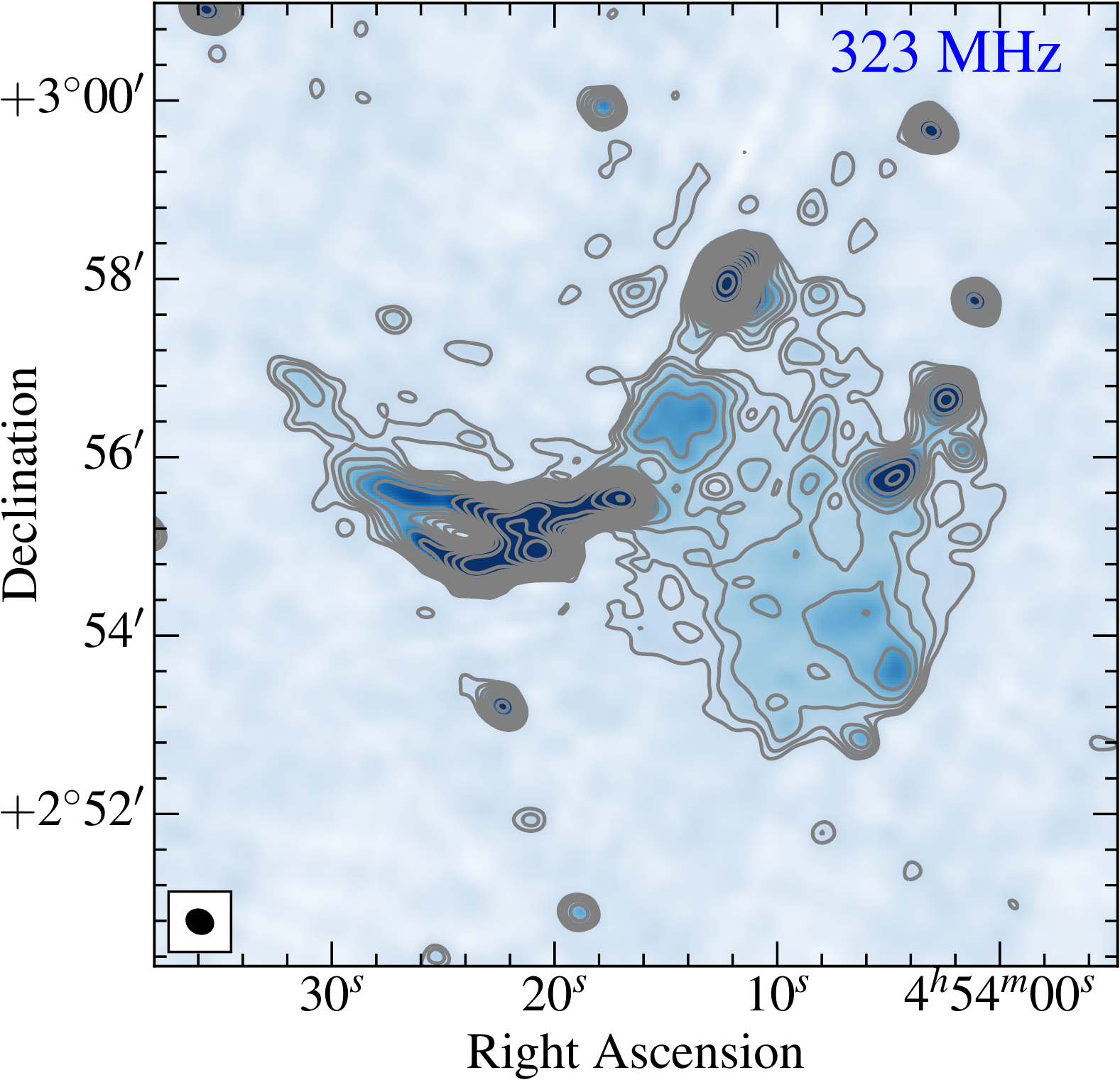}\hfill
        \includegraphics[width=0.32\textwidth]{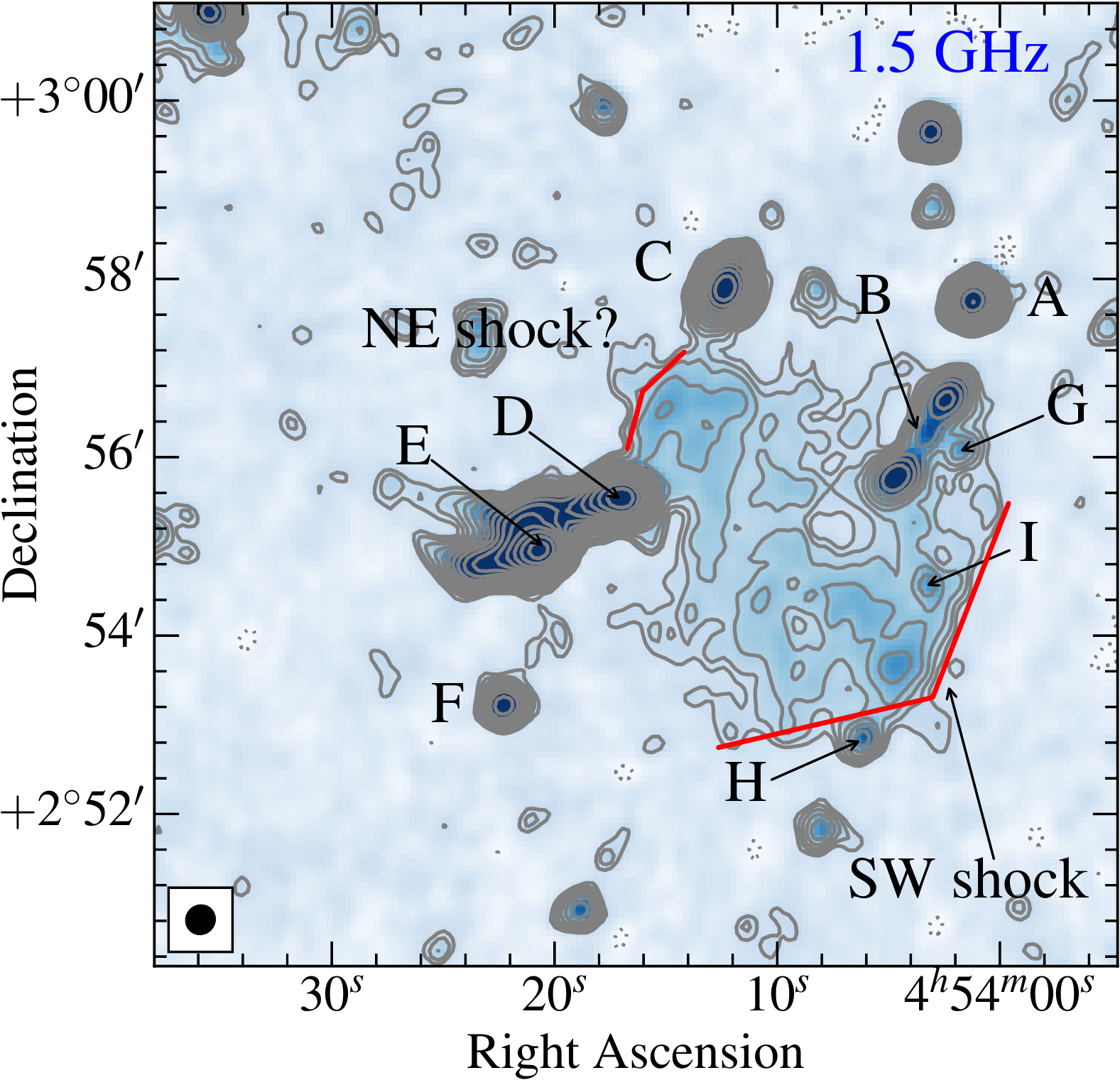}
        \caption{Radio continuum images of A520. The contours are from $\pm3\sigma$ and are spaced by $\sqrt{2}$. The image noise is $\sigma=460$, 100, and 26 $\upmu\text{Jy\,beam}^{-1}$ for the LOFAR, GMRT, and VLA images, respectively. The sources are labelled in the VLA image, partly adapting the notation in \cite{Vacca2014}. The synthesised beams are shown in the bottom left corners.}
        \label{fig:a520_mres}
\end{figure*}

\begin{figure*}
        \centering
        \includegraphics[width=0.49\textwidth]{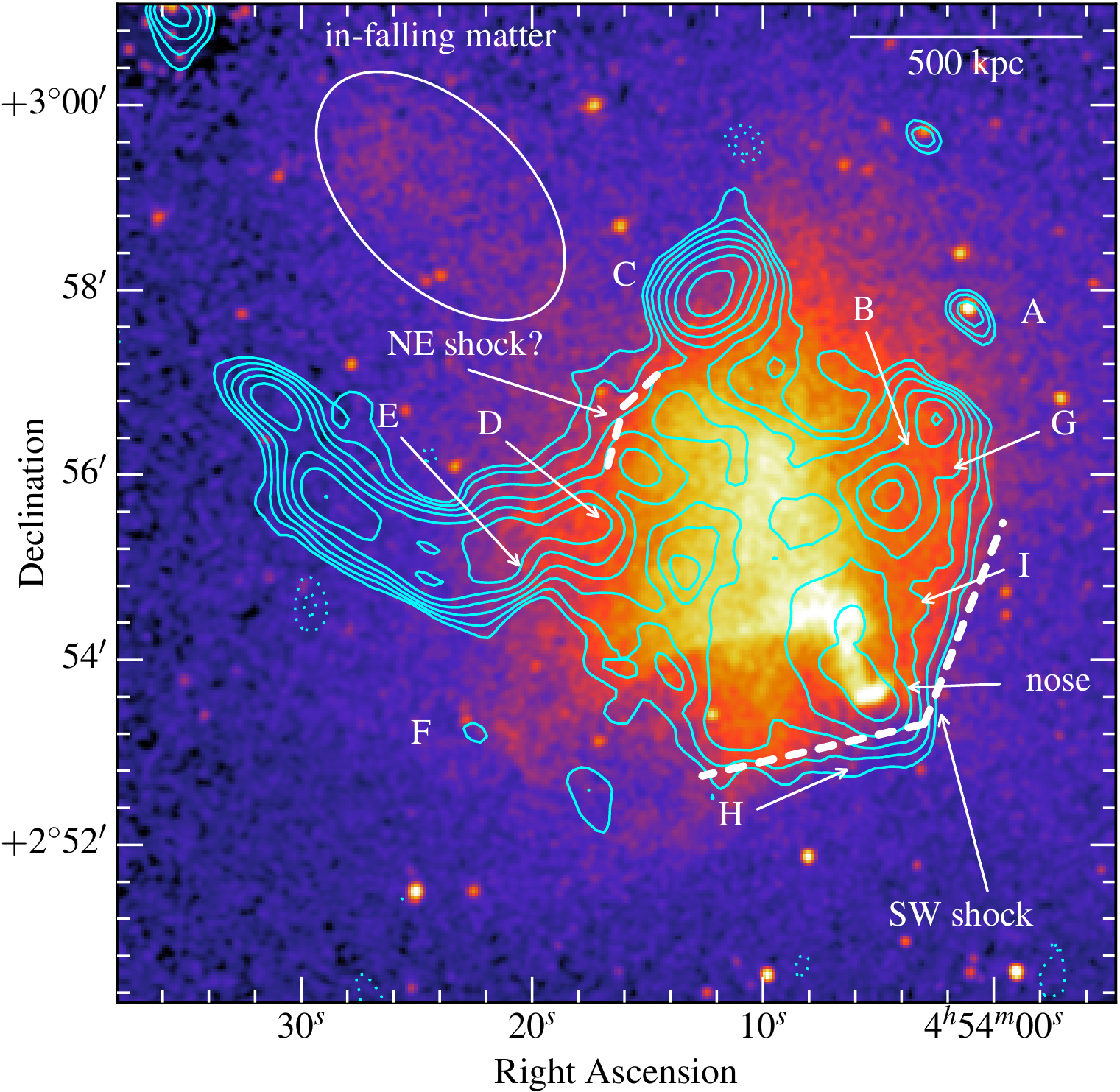}  \hfill
        \includegraphics[width=0.49\textwidth]{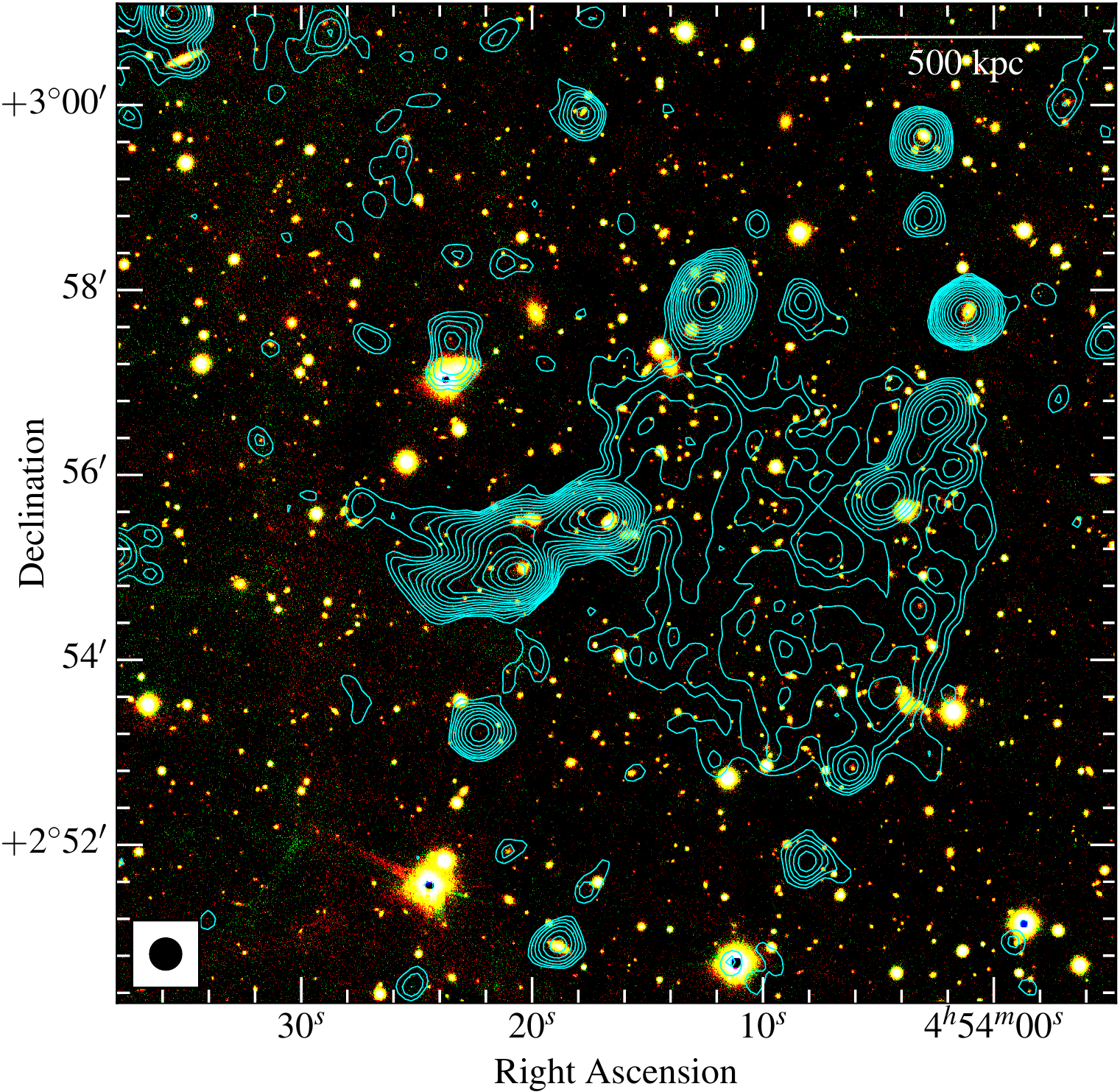} 
        \caption{\textit{Chandra} $0.5-2.0$ keV (left) and PanSTARRS colour (right) images of A520. The LOFAR (left) and GMRT (right) contours are identical to those in Fig. \ref{fig:a520_mres}.}
        \label{fig:a520_lofar_xray}
\end{figure*}

\begin{figure*}
        \centering
        \includegraphics[width=0.49\textwidth]{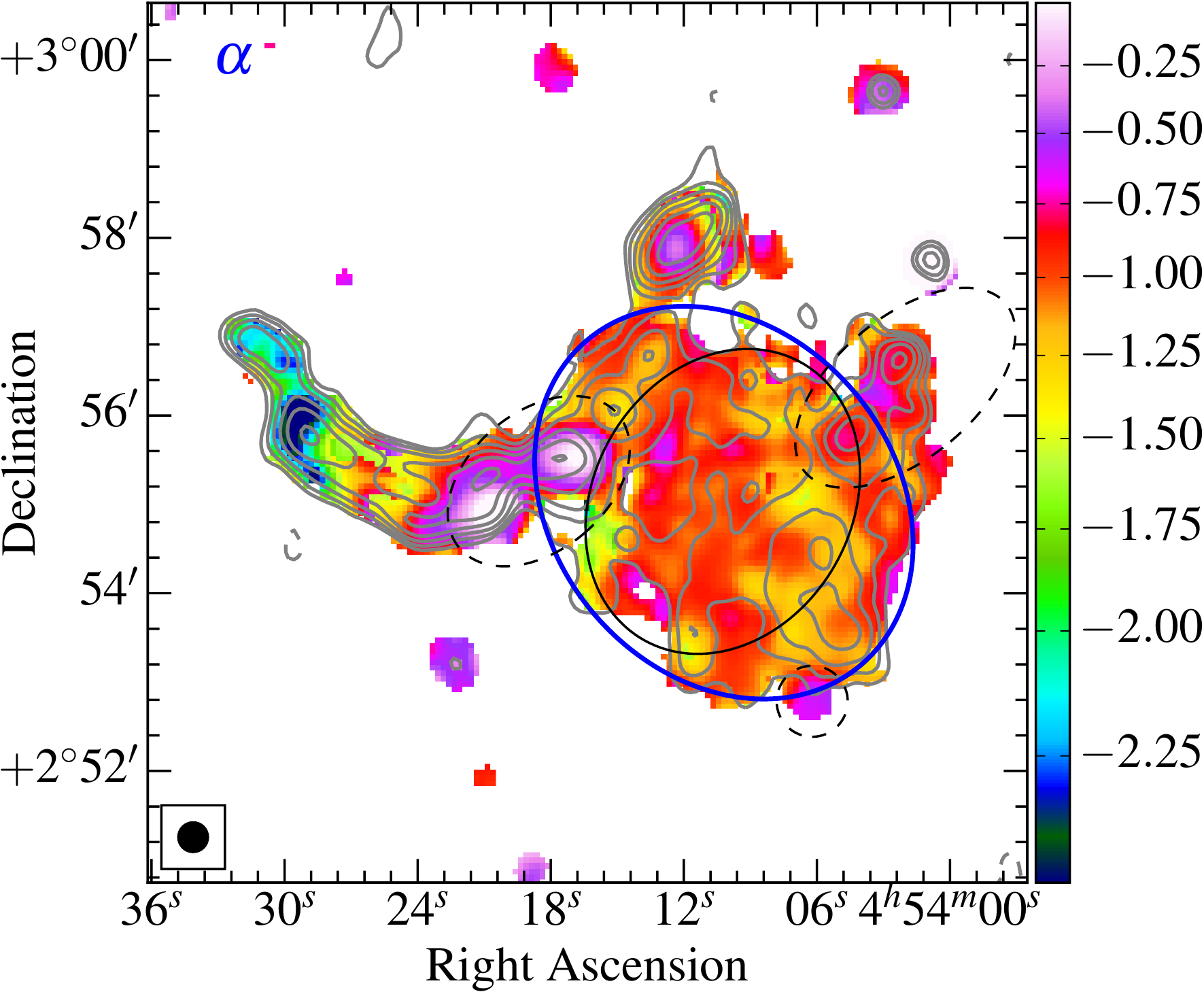} \hfill
        \includegraphics[width=0.475\textwidth]{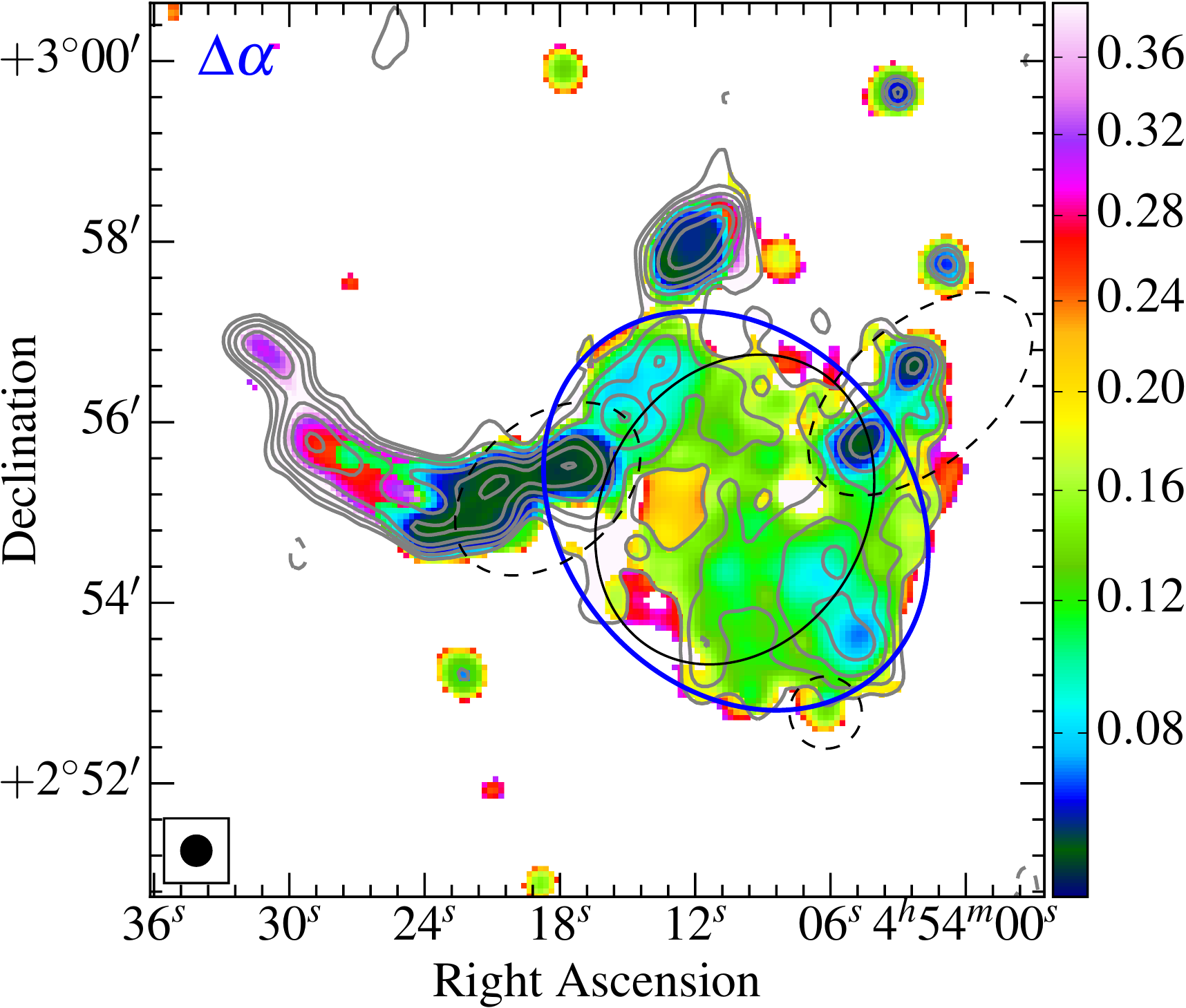}
        \caption{Spectral index from $145\,\text{MHz}$ to $1.5\,\text{GHz}$ (left) and error (right) maps of A520. In the both panels, the blue thick ellipse roughly follows the $3\sigma$ contour of the halo emission. The black thin ellipse shows the inner region of the halo. The black dashed regions are masked. The LOFAR contours begin with $3\sigma$, where $\sigma=450\,\upmu\text{Jy\,beam}^{-1}$, and are spaced with $\sqrt{2}$.}
        \label{fig:spx}
\end{figure*}

In Figs. \ref{fig:a520_mres} and \ref{fig:a520_lofar_xray}, we present the continuum images of A520 obtained with the LOFAR 145 MHz, GMRT 323 MHz, and VLA 1.5 GHz observations. The image properties are summarised in Table \ref{tab:image_para}. The radio images show the detection of diffuse emission associated with the ICM as well as multiple radio galaxies. To study the spectral properties of the extended radio sources, we combine the LOFAR, GMRT, and VLA data to make the spectral index map in Fig. \ref{fig:spx}.

\subsection{Radio halo}
\label{sec:res_halo}

Similar to previous observations, the new LOFAR 145 MHz and GMRT 323 MHz images (Fig. \ref{fig:a520_mres} and \ref{fig:a520_lofar_xray}) show the presence of a giant radio halo ($760\times950\,\text{kpc}^2$) with an overall morphology that approximately traces the \mbox{X-ray} emission of A520 \citep[e.g.][]{Giovannini1999,Govoni2001c,Vacca2014}. We used the LOFAR, GMRT, and VLA $20\arcsec$ maps that are made with identical imaging parameters (i.e. \textit{uv}-range $0.114-16.220\,k\lambda$, $\texttt{Briggs}$' $\texttt{robust}$ weighting 0.0) to measure the integrated flux for the halo emission. The measurement is done within an elliptical region (i.e. the blue thick ellipse in Fig. \ref{fig:spx}) with regions contaminated by the emission from the radio galaxies B, D, E, H, G, and I masked out. The elliptical region was chosen to roughly follow the $3\sigma$ contour of the LOFAR detected emission. Since the SB of the halo is approximately uniform within the $\lesssim450\,\text{kpc}$-radius central region (see Fig. \ref{fig:a520_mres}, also Fig. 5 in \citealt[][]{Vacca2014}), we assume that the SB in the masked regions can be extrapolated from the unmasked regions. The integrated flux of the halo is then calculated as the integrated flux measured in the unmasked region multiplied by a factor to account for the area of the masked regions (see Fig. \ref{fig:spx}). The total error quoted for the integrated flux measurements is the quadratic sum of the uncertainty in the flux scale and image noise. The integrated fluxes of the halo at 145 MHz, 323 MHz, and 1.5 GHz are $229.7\pm34.8$ mJy, $90.5\pm9.2$ mJy, and $18.8\pm1.0$ mJy, respectively. For comparison with the \cite{Wang2018a} measurement, which uses the same VLA data, we repeat the calculation for the flux within a region encompassing the $1\sigma$ contour. The halo flux within this region is $20.6\pm1.1\,\text{mJy}$ and is consistent with the value of $20.2\pm1.5\,\text{mJy}$ that was measured in a radio galaxy subtracted image presented in \cite{Wang2018a}. Another flux measurement for the extended emission at 1.4 GHz in \cite{Cassano2013a} is $19.4\pm1.4$ mJy, which is in agreement with our estimate. \cite{Vacca2014} obtain a lower value of $16.7\pm0.6$ mJy for the halo flux, but they did not extrapolate their measurements to estimate the contribution from masked regions, which might explain the difference.

\begin{figure}
        \centering
        \includegraphics[width=0.48\textwidth]{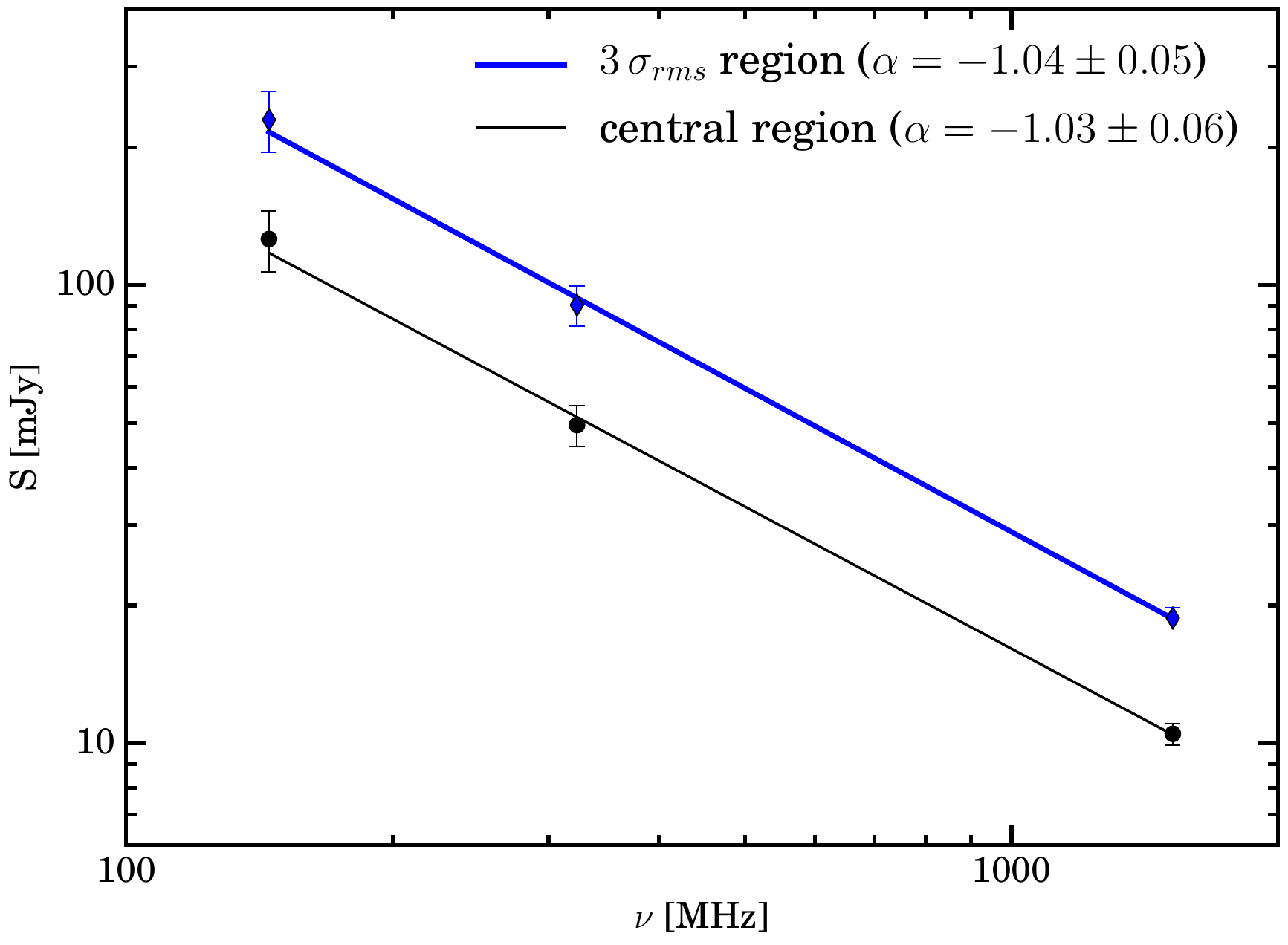}
        \caption{Integrated spectrum between 145 MHz and 1.5 GHz for the radio halo. The integrated fluxes are measured within the $3\sigma$ contour (blue thick) and central (black thin) regions in Fig. \ref{fig:spx}.}
        \label{fig:spx_int}
\end{figure}

To estimate the spectral index of the halo, we fit the integrated fluxes with a single power-law function, $S\propto\nu^\alpha$, which describes the synchrotron emission mechanism. As plotted in Fig. \ref{fig:spx_int}, the integrated spectrum of the halo follows the power-law relation with an index of $\alpha_\text{145 MHz}^\text{1.5 GHz}=-1.04\pm0.05$. Our measurement of the integrated spectral index is in line with the estimate of $\alpha_\text{325 MHz}^\text{1.4 GHz}=-1.12\pm0.05$ in \cite{Vacca2014}. In the central region of the halo, the integrated spectral index is estimated to be $\alpha_\text{145 MHz}^\text{1.5 GHz}=-1.03\pm0.06$. Examining the spatial distribution of the halo spectrum, we find that the spectral index over the inner region of the halo (i.e. black thin ellipse in Fig. \ref{fig:spx}) remains approximately constant around a mean of $-1.01$ with a scatter of $0.12$. A similar situation is found in the larger region (i.e. $\overline{\alpha}=-1.03$ in the blue thick ellipse in Fig. \ref{fig:spx}). However, Fig. \ref{fig:spx} also indicates that there are small changes of the spectral indices in the NE and SW region on scales larger than the beam size. These estimates of the spectral index, based on the distribution of the spectral indices, are also consistent with the integrated spectral index values above.

\subsection{Southwest region of the radio halo}
\label{sec:res_sw}

\begin{figure}
        \centering
        \includegraphics[width=0.48\textwidth]{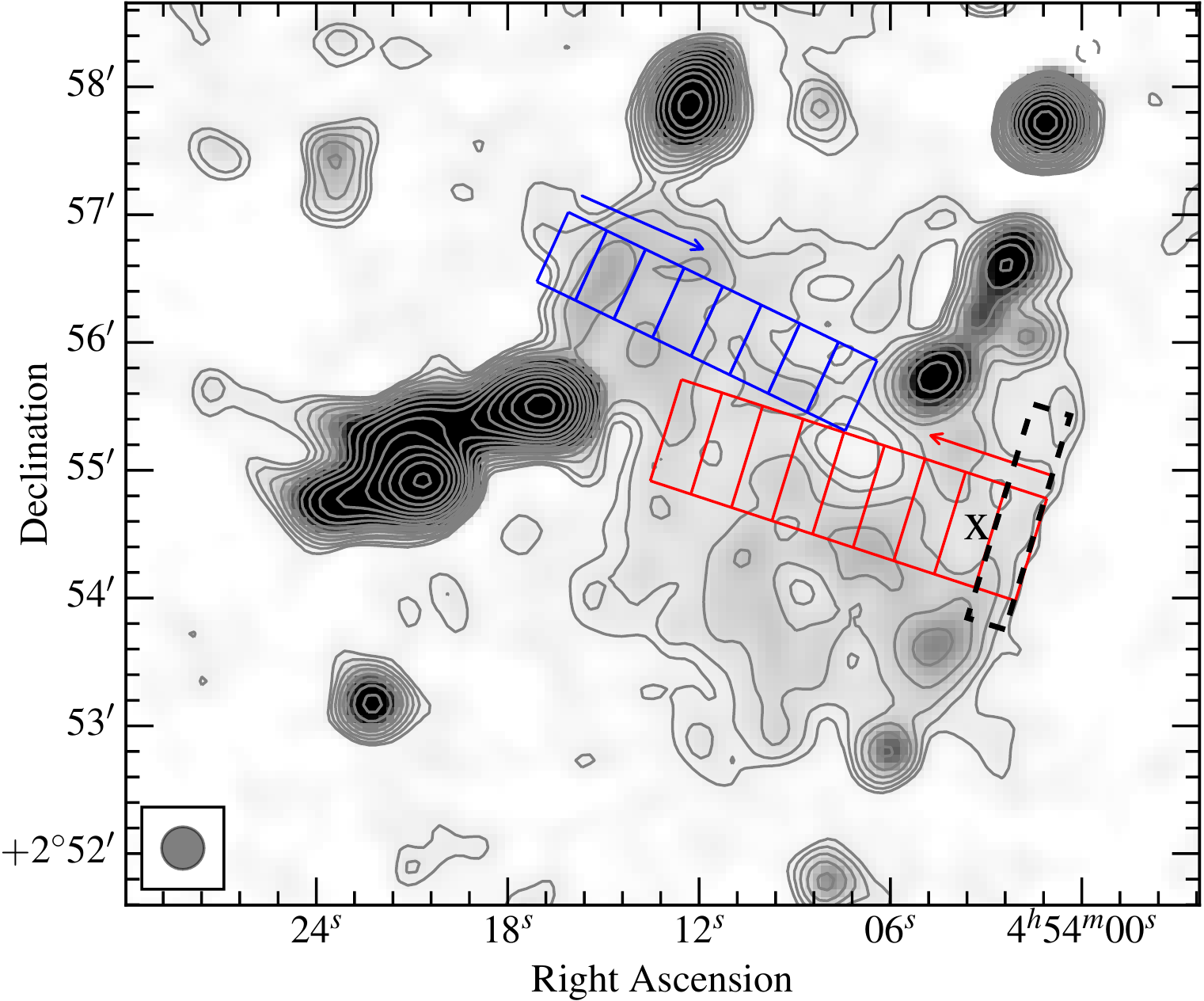}
        \caption{Rectangular regions in the SW and NE directions where SB and spectral indices are plotted in Fig. \ref{fig:sb_spx_pro}. The black dashed rectangle is where the spectral index right behind the SW shock is estimated. The X cross indicates the location of a compact source I that is subtracted from the VLA data. The VLA contours are identical to those in Fig. \ref{fig:a520_mres} (right).}
        \label{fig:sb_spx_pro_regions}
\end{figure}

\begin{figure}
        \centering
        \includegraphics[width=0.4\textwidth]{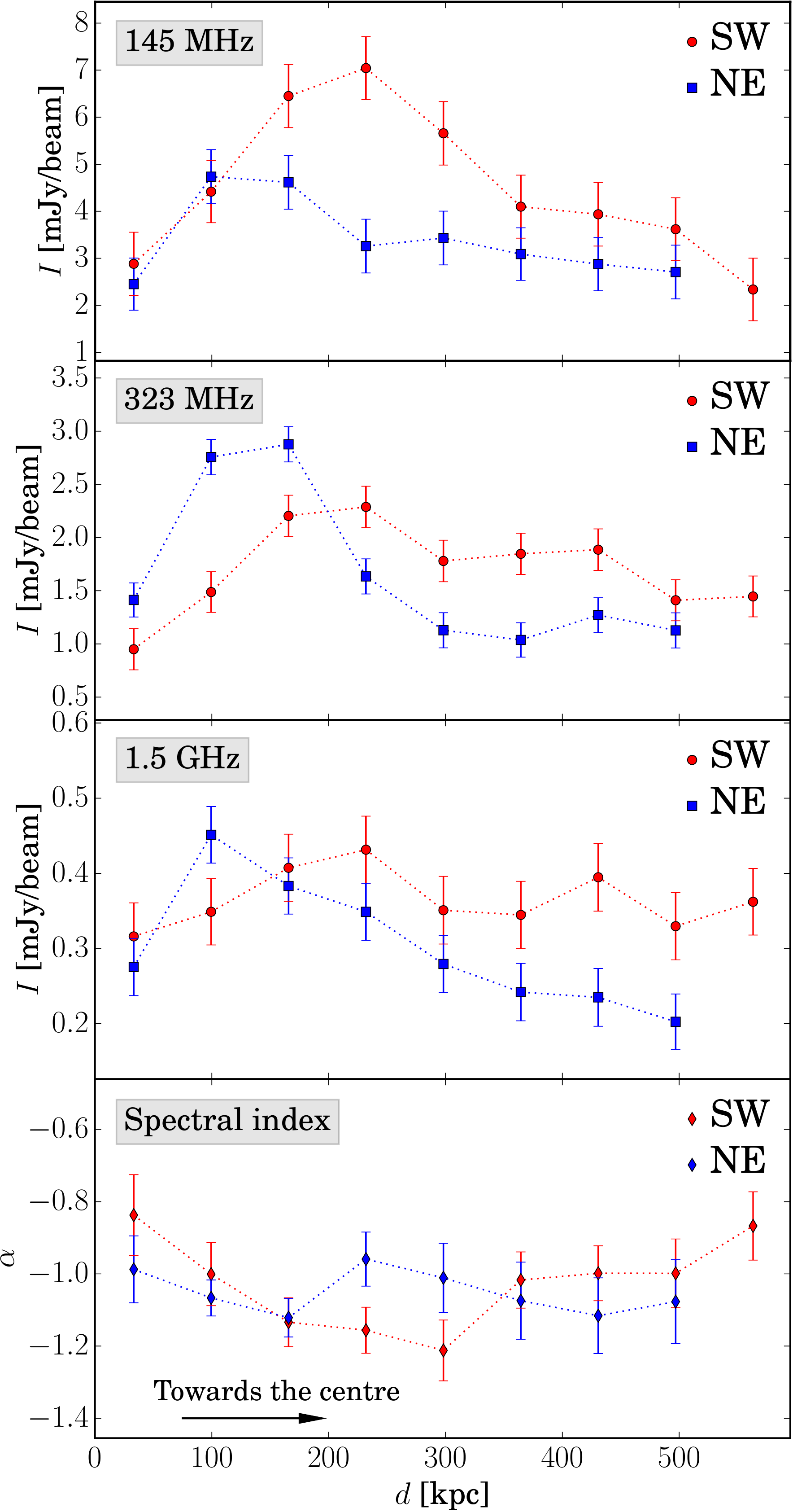}
        \caption{Surface brightness (top panels) and spectral index (bottom panel) profiles from outer edges towards cluster centre for the regions in the SW and NE directions in Fig. \ref{fig:sb_spx_pro_regions}. Towards the cluster centre, the spectral index is steepened for the SW radio emission and remains approximately constant for the NE radio emission. 
        }
        \label{fig:sb_spx_pro}
\end{figure}

The radio continuum images in Fig. \ref{fig:a520_mres} show excess emission in the SW region of the radio halo, which is consistent with previous observations  in for example \cite{Giovannini1999}, \cite{Govoni2001c}, and \cite{Vacca2014}. The SW radio emission roughly follows the bullet-like shock front detected with the \mbox{X-ray} observations \citep{Markevitch2005}. The radio emission increases sharply across the SW shock front from west to east. Assuming that the upper limit for the radio emission in the pre-shock region is $1\sigma$ of the background noise, the increase in the radio SB is $\sim4$, $\sim5$, and $\sim8$ times at 145 MHz, 323 MHz, and 1.5 GHz, respectively. In Fig. \ref{fig:sb_spx_pro_regions}, we plot the regions where the SB and spectral indices are extracted to examine the spatial distribution of the radio emission and spectral energy distribution. The profiles in Fig. \ref{fig:sb_spx_pro} show that the radio emission in the region behind the SW shock continues to increase, most significantly at low frequencies, before gradually decreasing in the region $\sim230\,\text{kpc}$ away from the SW radio edge. The spectrum index in the SW radio edge is flattest at the $3\sigma$ SW edge with  $-0.84\pm0.11$ and steepens to $-1.13\pm0.07$ at $\sim165\,\text{kpc}$  towards the cluster centre from the SW $3\sigma$ contour. This spectral trend can also be seen in the spectral index map (Fig. \ref{fig:spx}). Our spectral index measurement between 145 MHz and 1.5 GHz at the SW radio edge is flatter than the previous estimate of $\alpha_\text{323\,MHz}^\text{1.4\,GHz}=-1.25,$ which was derived from lower ($39\arcsec$) resolution data in \cite{Vacca2014}.

\subsection{Northeast region of the radio halo}
\label{sec:res_ne}

\begin{figure*}
        \centering
        \includegraphics[width=0.469\textwidth]{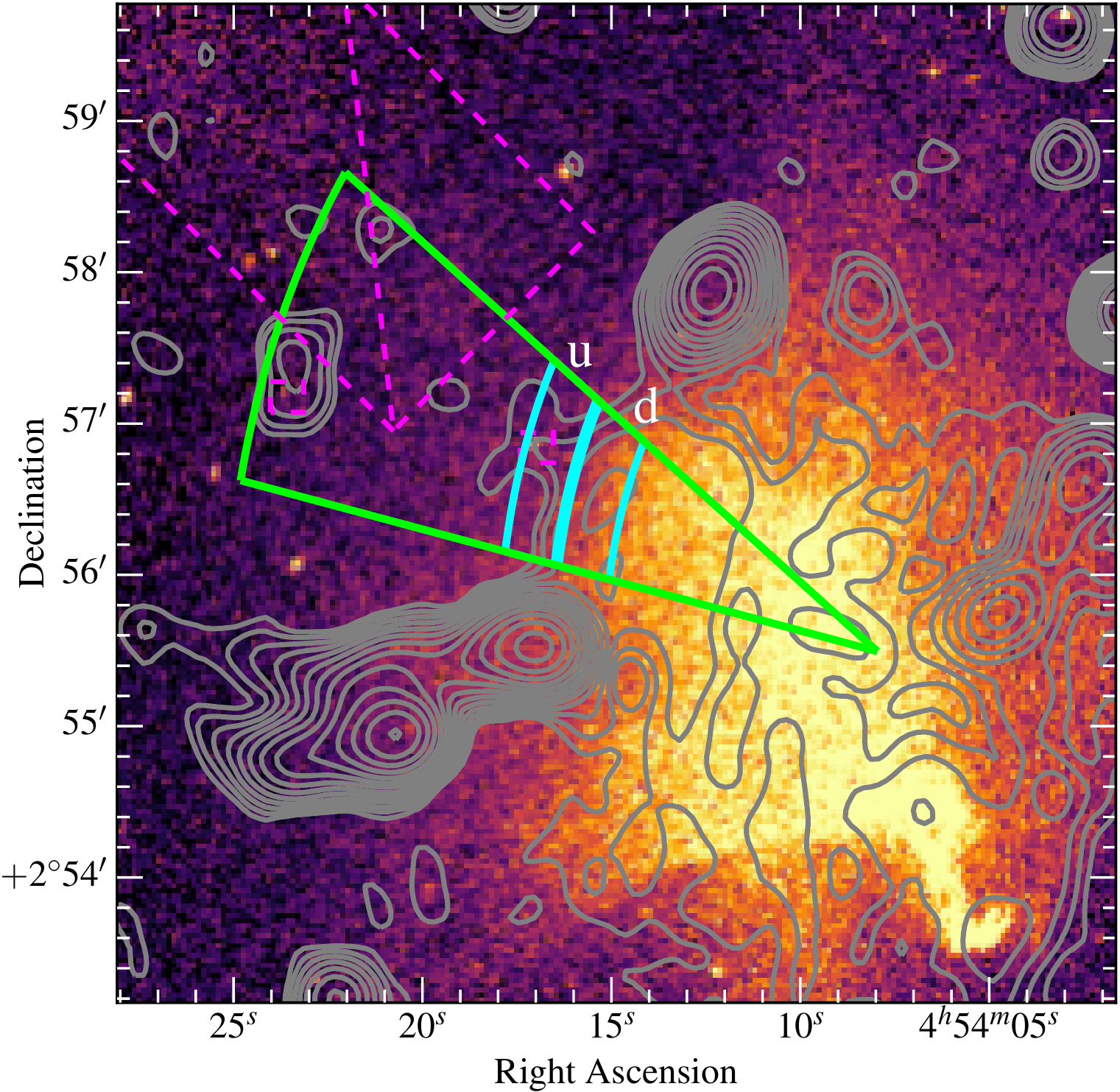} \hfill
        \includegraphics[width=0.48\textwidth]{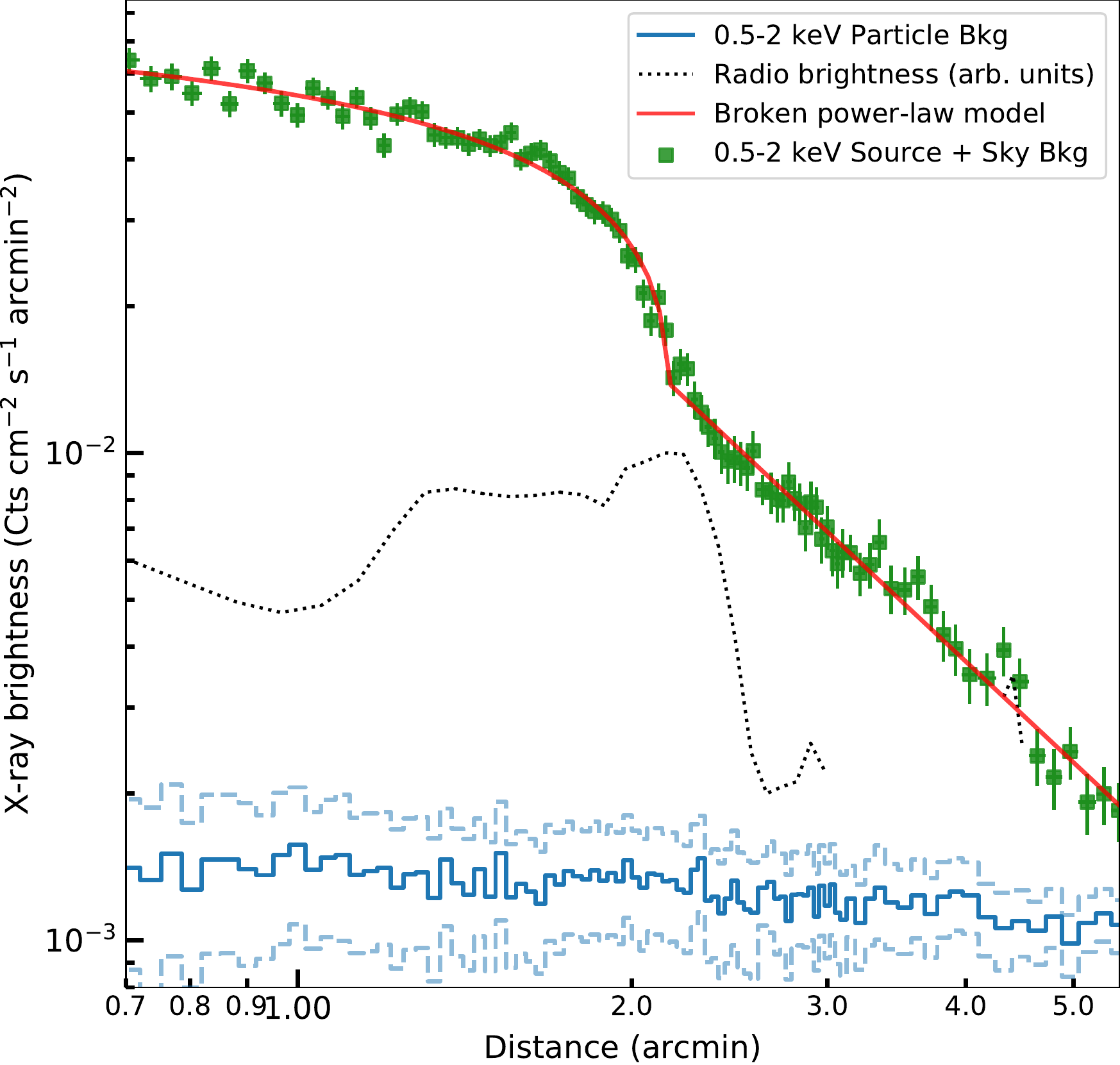}
        \caption{Left: the sector where SB are extracted from the \textit{Chandra} \mbox{X-ray} data. The magenta dashed regions are masked to remove the point sources and the emission from the NE plasma that is falling into the cluster. The blue regions represent where temperature is calculated for upstream and downstream ICM regions. Right: SB profile towards the NE region and the best-fit radial SB profile (red).}
        \label{fig:x_profile}
\end{figure*}

Towards the NE direction, the radio emission of the halo gradually increases before sharply decreasing at the NE edge (Fig. \ref{fig:a520_mres}). An enhancement in the extended radio emission (i.e. projected size of $\sim65\,\text{kpc}$ in radius) is detected at $\sim3\sigma$ at 1.5 GHz, but is slightly less visible at the lower frequencies of 145 MHz and 323 MHz. At the NE edge there is a hint that the spectral index seems to steepen in the NE-SW direction, but the steepening trend of the spectral index is still uncertain owing to the large errors in the spectral measurement (see the spectral index profile in Fig. \ref{fig:sb_spx_pro}, bottom panel).

A number of merging clusters generate double shocks that propagate on diametrically opposite sides of the cluster centre  \citep[e.g.][]{Russell2011,Dasadia2016,Urdampilleta2018}. In A520, a shock ($\mathcal{M}_{X}=2.1$; \citealt{Markevitch2005}) is detected in the SW region of the cluster and \cite{Wang2016} discussed a possibility of a counter shock in the NE side of the cluster. To further search for a possible counter shock, we fit the \mbox{X-ray} SB emission with a broken power-law function\footnote{using $\texttt{PyXel}$, available on \url{https://github.com/gogrean/PyXel}}, assuming that the \mbox{X-ray} emission is spherically symmetric about a point at the cluster centre. The fitted function is consistent with underlying broken power-law particle density profile,
\begin{equation}
n(r) =  \left\{
\begin{array}{lr}
C\,n_0\,\left(\frac{r}{r_\text{break}}\right)^{-a_\text{d}} & \text{if } r\leq r_\text{break} \\
n_0\left(\frac{r}{r_\text{break}}\right)^{-a_\text{u}}      & \text{if } r>r_\text{break}
\end{array}
\right.
\label{eq:x_profile}
,\end{equation}
where $C=n_\text{d}/n_\text{u}$ is the shock compression factor or particle density jump (the subscripts $d$ and $u$ stand for downstream and upstream regions, respectively); $r_\text{break}$ is the location of the SB discontinuity; $a$ is the slope of the power-law function; and $n_0$ is normalisation factor or particle density at the $r_\text{break}$ location of the upstream region. We find the best-fit parameters for the selected sector (see Fig. \ref{fig:x_profile}) that has a compression factor of $C=1.74\pm0.07$, coincidentally close to the outer edges of the radio emission. The SB jump is located $428.6^{+1.2}_{-0.9}\,\text{kpc}$ from the sector centre (i.e. $4^\text{h}54^\text{m}07.96^\text{s},+2^\text{d}55^\text{m}29.64^\text{s}$). The best-fit indices are $a_\text{d}=0.23\pm0.02$ and $a_\text{u}=1.56\pm0.17$. The normalisation factor is $n_\text{0}=(20.1\pm0.4)\times10^{-3}\,\text{cm}^{-1}$. The 90 percent uncertainties of the best-fit parameters have been taken into account the uncertainty of 3 percent in the quiescent background as used in \cite{Wang2016}. The particle density jump suggests the presence of either a cold front or shock front at the $r_\text{break}$ location. 

If the detected discontinuity in the \mbox{X-ray} SB is due to a cold front, the ICM thermal pressure must be balanced in the upstream and downstream regions (or  $T_\text{u}/T_\text{d}=C$). To check this possibility, we estimate temperature on both sides of the density jump location from the \textit{Chandra} data. The selected upstream and downstream regions are shown in Fig. \ref{fig:x_profile}. We obtain a temperature of $T_\text{u}=8.57_{-1.16}^{+1.48}\,\text{keV}$ for the pre-shock region and a slightly higher temperature in the post-shock region $T_\text{d}=9.49_{-0.88}^{+0.90}\,\text{keV}$, but this is still within the $1\sigma$ significance. \cite{Wang2016} estimated a slight decrease in the temperature in the post-shock region (i.e. $\sim9\,\text{keV}$), but the decrease is still within their $1\sigma$ uncertainty and they use larger regions (i.e. C3 and C4 in their Fig. 4) than the regions we used in Fig. \ref{fig:x_profile}  in this
work. It is also noted that the temperature estimate in the NE region might be contaminated by other sources such as the in-falling hot matter (see Fig. \ref{fig:a520_lofar_xray}) or the nearby \mbox{X-ray} emission in the regions of sources C, D, and E \citep{Wang2016}. The temperature value we estimate implies that if the SB jump is due to a cold front then $C=0.90\pm0.16$. This is inconsistent with the SB jump of $C=1.74\pm0.07$ from the broken power-law fitting of the \mbox{X-ray} emission. Therefore, the assumption that the SB jump is because of the presence of a cold front might be unlikely. 

In the case in which the discontinuity is caused by a merger shock front, the compression of the ICM plasma is directly related to the shock Mach number by the Rankine-Hugoniot jump relation,
\begin{equation}
        \mathcal{M}_{X}=\sqrt{\frac{2C}{\gamma+1-C(\gamma-1)}},
\end{equation}
where $\gamma=5/3$ is the adiabatic index of the ICM plasma. Given the compression factor of $C=1.74\pm0.07$, we have $\mathcal{M}_\text{NE}^{X}=1.52\pm0.05,$ which is smaller than the SW shock Mach number ($\mathcal{M}_\text{SW}^{X}=2.4_{-0.2}^{+0.4}$; \citealt{Wang2018a}), but is still in the range of values that are typically  estimated for cluster merger shocks (i.e. $\mathcal{M}_{X}\lesssim3$) in the literature \citep[e.g.][]{Russell2010,Macario2011a,Akamatsu2015,Botteon2016a,Botteon2016b,Dasadia2016}. The shock Mach number can also be related to the temperature jump at the shock,
\begin{equation}
        \mathcal{M}_{X}=\sqrt{\frac{[(8T_{du}-7)+[(8T_{du}-7)^{2}+15]^{1/2}}{5}},
\end{equation}
where $T_{du}=\frac{T_d}{T_u}$ \citep[e.g.][]{Landau1959,Markevitch2007,Finoguenov2010,VanWeeren2016b}. The Mach number derived from the temperature is $\mathcal{M}_\text{NE}^{X}=1.1_{-0.2}^{+0.3}$, which is slightly smaller than the value we estimate from the SB jump above. As mentioned, the precise temperature value might be biased by the contaminating sources in the NE region, which are not be easily removed \citep[e.g.][]{Wang2016}. 

\section{Discussion}
\label{sec:dis}

\subsection{Radio halo}
\label{sec:dis_halo}

\begin{figure*}
        \centering      
        \noindent\begin{minipage}[t]{0.277\textwidth}
                \centering
                \begin{tikzpicture}
                \node[anchor=south west,inner sep=0] (image) at (0,0) {\includegraphics[width=1\columnwidth]{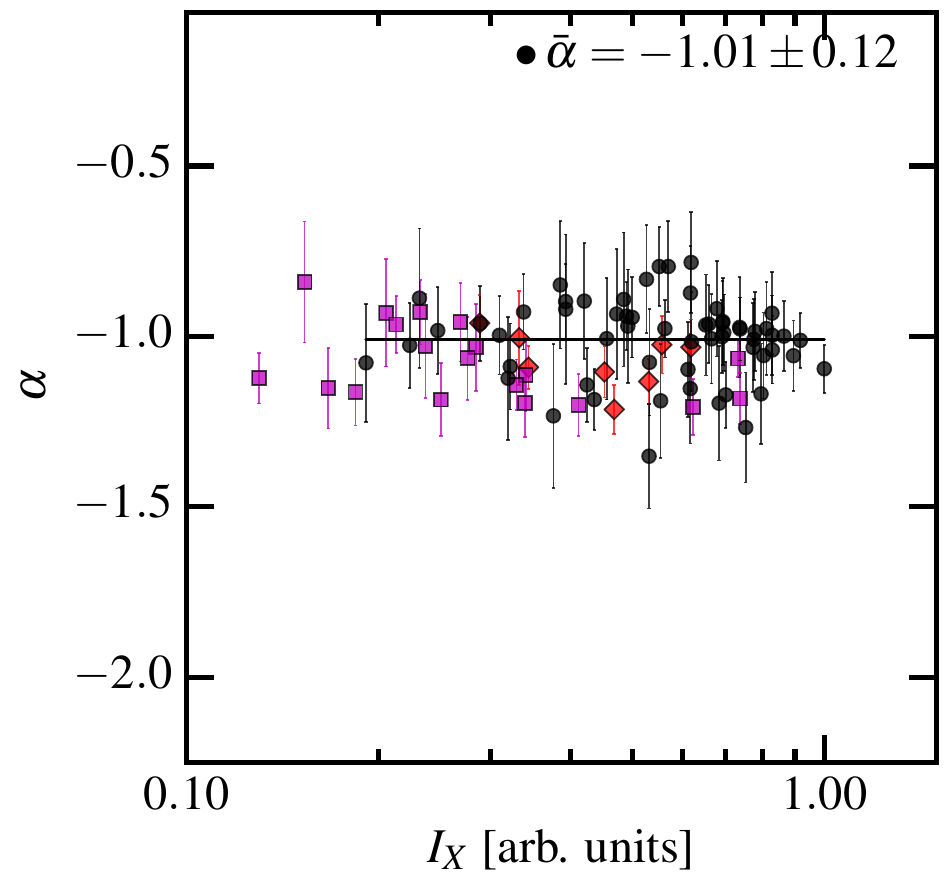}};
                \begin{scope}[x={(image.south east)},y={(image.north west)}]
                \node[anchor=south west, inner sep=0] (image) at (0.22,0.18) {\includegraphics[width=0.25\columnwidth]{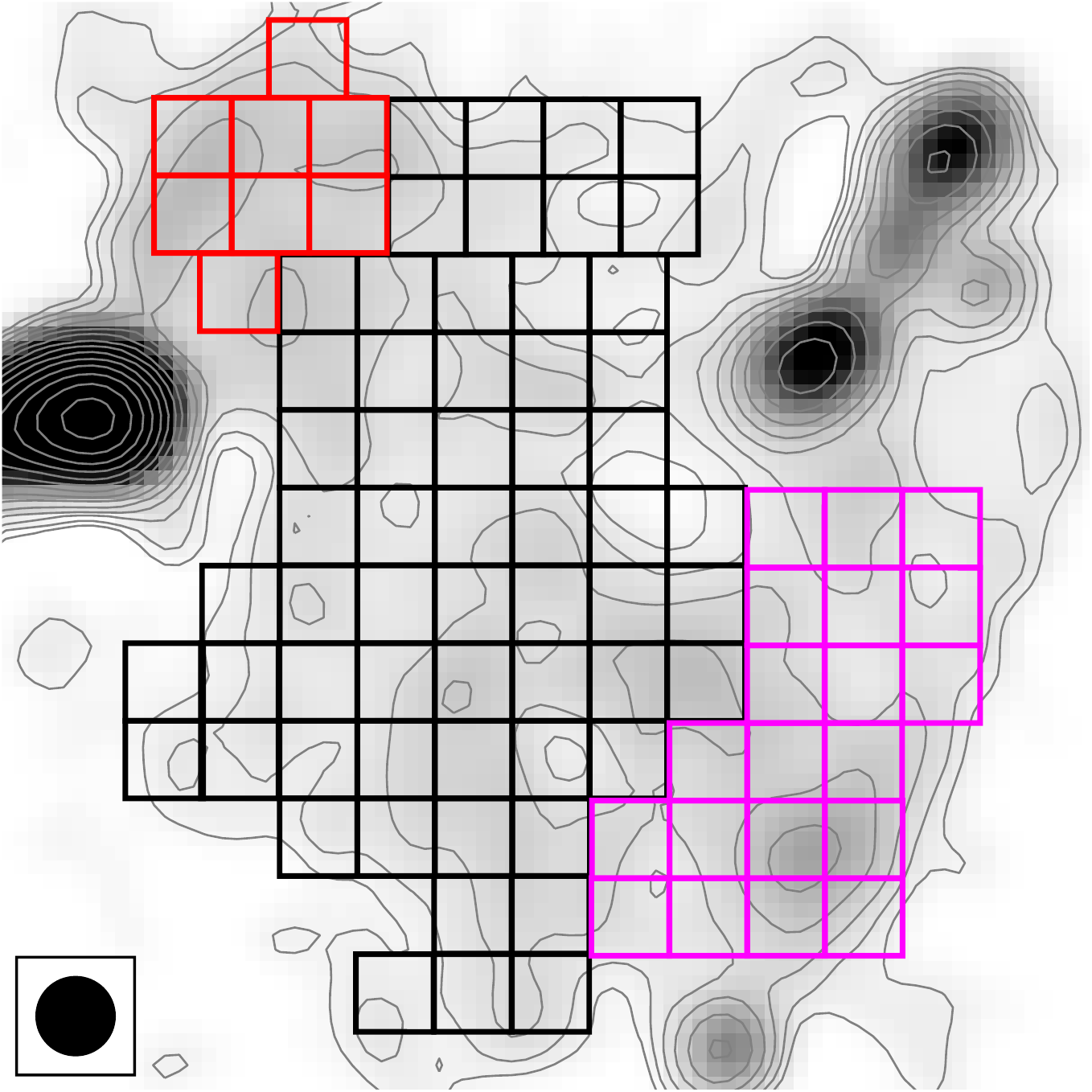}};
                \end{scope}
                \end{tikzpicture}
        \end{minipage}
        \includegraphics[width=0.713\textwidth]{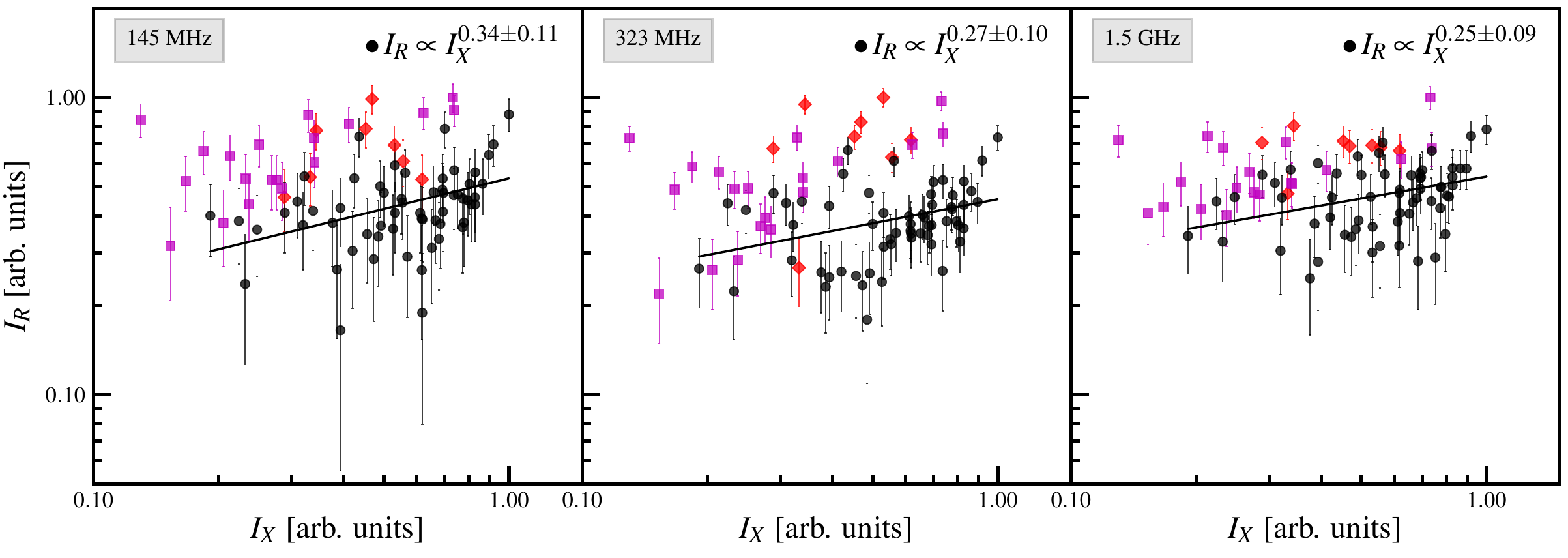}
        \caption{Scatter plots of spectral index and radio SB ($I_R$) as a function of \mbox{X-ray} SB ($I_X$). The regions for extracting data are shown in the overlaid image (left). The data points are plotted in the same colour as the regions in the overlaid image (i.e. black circles for the central regions, magenta squares for the regions in the SW direction, and red diamonds for the regions in the NE direction). The best-fit parameters are obtained for the central regions.}
        \label{fig:r_x_correlation}
\end{figure*}

Despite the accurate characterisation of the radio emission in A520, it remains uncertain whether or not the emission can be entirely attributed to a halo or if there is additional radio relic emission  \cite[e.g.][]{Govoni2001c,Vacca2014}. The radio halo coincides with the region swept by the SW shock and by the (possible) NE shock.
Shocks are detected coincident with the edges of a number of radio haloes \citep[e.g.][]{Markevitch2005,Markevitch2010,Macario2011a,Shimwell2014}, suggesting that they might drive a turbulent region bounded by tangential magnetic fields in which particles are confined and \mbox{re-accelerated}.

Although the overall shape of the halo follows the \mbox{X-ray} emission, the radio brightness of the halo is very flat, especially at 1.5 GHz (Fig. \ref{fig:a520_mres}). A correlation between radio and X-rays is observed only in the region of the trail/bullet-like structure in the southern region where the radio and \mbox{X-ray} emission appears brighter (Fig. \ref{fig:a520_lofar_xray}). However, in general the radio brightness does not follow the \mbox{X-ray} emission. This is clear in Fig. \ref{fig:r_x_correlation} where we show the point-to-point radio to \mbox{X-ray} brightness using cells of the beam size (i.e. $20\arcsec$ or $66\,\text{kpc}$). Basically only a tentative trend in the form of $I_R\propto I_X^b$, where $b=0.34\pm0.11$, $0.27\pm0.10$, and $0.25\pm0.09$ at 145 MHz, 323 MHz, and 1.5 GHz, respectively, is observed in the central region of the halo. This is similar to a case reported by \cite{Shimwell2014} who found no significant correlation between radio halo and \mbox{X-ray} brightness in the bullet cluster \mbox{1E 0657$-$55.8}. However, the relation between the radio and \mbox{X-ray} brightness in A520 is different from a number of cases of regular and roundish radio haloes reported in the literature where a clear correlation, typically slightly sublinear, is found between the radio and \mbox{X-ray} brightness (e.g. $b\approx0.64-0.99$; \citealt{Govoni2001a,Govoni2001c,Feretti2001,Venturi2013,Rajpurohit2018}). This suggests that the haloes in A520 and the bullet cluster may be at similar evolution states, but their states may be different from those of the haloes in the clusters that follow the scaling relation.

All models that have been proposed for the formation of radio haloes predict a connection between radio and \mbox{X-ray} emission (e.g. see \citealt{Brunetti2014} for review). This is straightforward in hadronic models (including the models where secondary particles are \mbox{re-accelerated} by turbulence) because the ICM that produces the \mbox{X-ray} radiation also provides the targets for the inelastic collisions that generate the radio emitting secondary electrons. Also in the case in which primary seed electrons are \mbox{re-accelerated} by turbulence, the energy reservoir to \mbox{re-accelerate} particles is extracted from the turbulent and kinetic energy of the thermal ICM. This should induce some connection between thermal and non-thermal quantities. One possibility to explain the properties of the extended radio emission in A520 is that the radio and \mbox{X-ray} emission are not co-spatial and that the majority of the radio emission is produced in a thick perturbed cocoon that bounds the volume. The halo is swept by the outgoing shocks and embeds the very central dense substructures that appear bright in the \mbox{X-ray} band. This situation may explain the connection between the edges of the halo and shocks and, in fact, it is not surprising given the very particular evolutionary stage of A520. According to \cite{Wang2016}, the very peculiar twisted structure that dominates the \mbox{X-ray} emission in A520 traces the gas from a disrupted cool core that is swept away from the central galaxy of its former host sub-cluster by ram pressure in the downstream region of the SW shock. This gas is observed to form a physically connected trail of dense and cold clumps resembling a leg with a bright foot, knee, and leg along the ridge extending about 300 kpc east from the knee \citep{Wang2018a}. The twisted structure suggests that motions in the shocked gas surrounding the trail are turbulent, however the cool gas within the trail itself is probably not turbulent because at this early stage ram pressure stripping prevents the development of instabilities at its boundaries and the mixing of the trail with the hot surrounding medium \citep{Takizawa2005}.

Another important clue to the origin of the radio halo is provided by the spectral index distribution in A520. As already mentioned in Sec. \ref{sec:res_halo}, the spectral index in the radio halo is fairly constant $\alpha \approx -1.03$ with a scatter of $\sim 0.12$ on beam scales (i.e. $20\arcsec$ or $66\,\text{kpc}$). We performed simulations to examine whether the scatter is due to statistical errors (i.e. from the image noise) or the intrinsic spectral features of the halo. First, we assume that the radio halo has a constant spectral index (i.e. $\alpha=-1.03$, meaning no intrinsic scatter). Using the observed VLA 1.5 GHz map, we generate the halo flux distribution at 145 MHz and 323 MHz. Gaussian noise is then added to the maps at levels that are estimated from the observed LOFAR, GMRT and VLA maps (see Table \ref{tab:image_para}). The spectral index maps derived from the simulated radio maps are used to calculate the anticipated scatter in spectral index. The cells we used have the size of the beam (i.e. $20\arcsec$; see Fig. \ref{fig:r_x_correlation}). We find that the observed spectral index scatter of $0.12$ on the scales of the beam size is consistent with the scatter recovered from the simulated spectral index maps (i.e. $\Delta\alpha_\text{sim.}=0.13\pm0.02$) where the statistical errors are from the image noise alone. On larger scales, there are hints of possible spectral variations, for example the southern region of the halo (the nose and bullet-like/trail; the magenta squares in Fig. \ref{fig:r_x_correlation}) shows a spectral index that is steeper than the SW region of the halo (Fig. \ref{fig:spx}), still spectral variations are fairly small, in the range $0.1-0.2$, if we exclude the regions in the SW direction (i.e. magenta squares). A more detailed investigation on spectral variations in the halo using, for example simulations and a theoretical analysis, is beyond the aim of this paper. However, the projected distribution of spectral indices in the halo can be used to infer basic constraints. For example, using a conservative limit $\Delta\alpha_\text{proj.}<0.12$ on spatial scales of the beam (i.e. $20\arcsec$) and assuming that the spectral index in the halo volume can change stochastically around a mean value, the intrinsic scatter is roughly $\Delta \alpha_\text{intr.} \approx \sqrt{N}\times\Delta\alpha_\text{proj.}$, where $N\approx10$ is the number of beam cells intercepted along the line of sight, implying $\Delta\alpha_\text{intr.} < 0.38$. Such a small-moderate scatter of the spectrum in the halo volume is also consistent with the hypothesis that the majority of the radio emission is generated in a smooth extended cocoon rather than from the central regions in which strong \mbox{X-ray} gradients and disrupted cool core are observed.

\subsection{South-west radio edge}
\label{sec:dis_sw_emission}

\cite{Markevitch2005} and \cite{Wang2018a} discussed whether the radio emission close to the prominent SW shock could be induced by adiabatic compression or \mbox{Fermi-I} processes. The \mbox{Fermi-I} processes generate synchrotron emitting relativistic electrons by accelerating electrons either directly from the thermal pool or \mbox{re-accelerating} mildly relativistic electrons from a seed population of pre-existing fossil plasma. In this section we reassess these possibilities making use of our new measurements.

\subsubsection{Shock acceleration}
\label{sec:dis_dsa}

In the DSA model, relativistic electrons with the Lorentz factor \mbox{$\gamma\gg10^{3}$}, which emit radio synchrotron emission in $\sim\upmu\text{G}$ magnetic fields, are \mbox{(re-)accelerated} by a shock. The sources of the relativistic electrons could be either the thermal electrons or a pre-existing population of fossil electrons in the ICM. Unless the fossil electrons have flat spectrum, the spectral index ($\alpha_\text{inj}$) of the injected CR electrons is related to the shock Mach number $\mathcal{M}$ \cite[e.g.][]{Blandford1987},
\begin{equation}
\alpha_\text{inj} = \frac{1}{2}-\frac{\mathcal{M}^2+1}{\mathcal{M}^2-1}.
\label{eq:M_inj}
\end{equation}
The relativistic electrons injected by the shock have an energy spectrum distribution of $\frac{dN}{dE}\propto E^{-\delta_\text{inj}}$, where $\delta_\text{inj}=1-2\alpha_\text{inj}$. In cases in which the electron cooling time is much quicker than the lifetime of the shock, the integrated spectral index of the radio emission in the region behind the shock is steeper than the injection index by 0.5 (i.e. $\alpha_\text{int}=\alpha_\text{inj}-0.5$; \citealt[][]{Ginzburg1969}).

The electron cooling due to the synchrotron and IC energy losses is one of the key observational links between large-scale shocks and extended radio emission in merging galaxy clusters. Observationally, the cooling results in a characteristic steepening of the spectral index of the radio emission with distance from the shock front \citep[e.g.][]{VanWeeren2010a}. Despite the SW region of A520 hosting a strong shock, previously radio observations have not detected the spectral steepening in the region behind the shock \citep{Vacca2014}. The lack of detection may be due to sensitivity and resolution limitations.

In Sec. \ref{sec:res_sw}, we used the $20\arcsec$-resolution images to show that the spectral index for the radio emission steepens in the region behind the SW shock front. The spectral index immediately behind the shock front is $\alpha_\text{145 MHz}^\text{1.5 GHz}=-0.85\pm0.06$. For a proper comparison with the \mbox{X-ray} study in \cite{Wang2018a}, we measure in this section the index in the post-shock region where the shock Mach number is highest (i.e. region N1+N2 in \citealt{Wang2018a}; also see the black dashed rectangle in Fig. \ref{fig:sb_spx_pro_regions}).  According to the DSA model, if the relativistic electrons in the SW edge are \mbox{(re-)accelerated} from the thermal pool or steep-spectrum fossil electrons with the injection index of $-0.85\pm0.06$, the shock should have a Mach number of $\mathcal{M}_\text{SW}=2.6^{+0.3}_{-0.2}$. Our estimate of the shock Mach number is in line with the measurements from \mbox{X-ray} data \citep[e.g.  $\mathcal{M}_\text{SW}^\text{X}=2.4_{-0.2}^{+0.4}$ in ][]{Wang2018a}. The agreement between the radio and \mbox{X-ray} derived Mach numbers for the SW shock implies that, in this case, the spectral properties of the radio emission at the SW edge are consistent with the DSA scenario in which the radio emitting relativistic electrons are either accelerated from the thermal pool or \mbox{re-accelerated} from a pre-existing population of fossil plasma.

In the shock-related \mbox{(re-)acceleration} scenario, radio emission is brightest at the shock front and becomes fainter in the downstream region where steeper spectrum emission should be detected \citep[e.g.][]{VanWeeren2010a,Rajpurohit2018}. However, the brightest emission in the SW region of A520 is found at a far distance (i.e. $\sim220\,\text{kpc}$) from the SW shock front and is located at the region of the steepest spectrum emission (Fig. \ref{fig:sb_spx_pro}), which seems to be inconsistent with the shock-related \mbox{(re-)acceleration}. A possibility for this mismatching is that the downstream shock-related emission is mixed with the emission from the halo due to smoothing and/or projection effects. Otherwise, the agreement between the radio and X-ray derived Mach numbers might be a random coincidence. To separate the two scenarios, polarisation observations will be required.

\subsubsection{Shock compression}
\label{sec:dis_compression}

Another possible mechanism for the radio emission at the SW edge is the gas compression of fossil electrons by the merger shock as  discussed in \cite{Markevitch2005}. Since magnetic fields are embedded in the ICM plasma, the adiabatic compression by the shock amplifies the magnetic field strength and increases the energy density of pre-existing relativistic electrons. A requirement for this scenario to happen is that both fossil electrons and magnetic fields must be present and produce synchrotron emission before the shock passage. Given a population of fossil electrons of a power-law energy spectrum, $dN/d\gamma=N_0\gamma^{-\delta}$, the synchrotron emissivity per unit volume immediately behind the shock front is proportional to the shock compression factor $C$,
\begin{equation}
        I_\nu\propto C^{(2/3)\delta+1}.
        \label{eq:I_post}
\end{equation}

Under the hypothesis of power-law distribution of the emitting electrons, the predicted relation, Eq. \ref{eq:I_post}, between the radio emission in the pre- and post-shock regions by the compression model provides means for observationally testing of the model. A recent study by \cite{Wang2018a} investigated this scenario using the VLA 1.4 GHz data, but no extended emission is detected in the pre-shock region. By modelling the radio emission in the shock regions, \cite{Wang2018a} have found that an emissivity jump of a factor of 10 (or 16) is inconsistent with the VLA data at $3\sigma$ (or $2\sigma$) significance. Given the estimated parameters for the shock (i.e. $\alpha=-1.25$, $C_\text{nose}=2.7$), \cite{Wang2018a} found a predicted emissivity jump of 16, meaning that the compression model is rejected at $2\sigma$ significance. If the CRs lose most of their energy within $130\,\text{kpc}$, the spectral index immediately behind the shock was approximated as $\alpha=-1.25+0.50=-0.75$. In this case, the compression model predicts a jump of 9, which increases the statistical significance of rejecting the compression model to $3\sigma$ \citep{Wang2018a}. The spectral index value used in \cite{Wang2018a} was estimated from a low-resolution ($39\arcsec$ or $130\,\text{kpc}$) spectral index map by \cite{Vacca2014}. This may bias the spectral measurement as high-resolution maps are required to resolve the spatial distribution of the source spectrum.

Our $20\arcsec$ resolution spectral index map in Fig. \ref{fig:spx} quantifies the steepening of the spectral index behind the SW shock front (also see the spectral index profile in Fig. \ref{fig:sb_spx_pro}, bottom). The spectral index between 145 MHz and 1.5 GHz in a $20\arcsec$ wide region behind the SW edge (i.e. the N1+N2 sector in \citealt{Wang2018a}; see Fig. \ref{fig:sb_spx_pro_regions}) is $\alpha=-0.85\pm0.06$. With this higher-resolution measurement of the spectral index, we estimate that the emissivity in the pre-shock region should be $10\pm3$ times lower than that in the region immediately behind the shock front, according to the adiabatic compression model (Eq. \ref{eq:I_post}). This prediction can be ruled out at the $2-3\sigma$ confidence levels, based on our new spectral index measurement and the modelling of the radio emission in the shock region in \cite{Wang2018a}. In the calculation, we have used a power-law spectrum slope of $\delta=1-2\alpha=2.64$ and a shock compression factor of $C(\theta)=C_\text{nose}\sqrt{\cos(\theta)}$, which is the correction for the azimuthal dependence in the measurement of the gas density jump (here $\theta$ is the angle from the shock nose; \citealt{Wang2018a}). We used  $C_\text{nose}=2.7\pm0.3,$ which is the maximum density jump at the location of the shock nose \citep[i.e. region N1 in ][]{Wang2018a}. The true spectral index at the SW shock front might be even flatter if the relativistic electrons radiate a significant amount of their energies within the $20\arcsec$ resolution region we use here. This would further lower the predicted emissivity jump and increase the significance at which this scenario could be ruled out.

The generation of the relativistic electrons at the SW shock is therefore unlikely to be solely caused by the adiabatic compression of the fossil electrons. However, this does not exclude the scenario in which the adiabatic compression and \mbox{re-acceleration} of fossil electrons by the shock simultaneously occur because the predicted emissivity jump, depending on the spectrum of the pre-existing relativistic electrons, might be considerably higher in this case \citep{Markevitch2005} and could still be consistent with the constraints obtained by \cite{Wang2018a} using the VLA data set.

\subsection{North-east radio edge}
\label{sec:dis_ne}

The origin of the excess radio emission in the NE region remains unclear. The detection of the SB \mbox{X-ray} jump coincident with the location of the NE radio edge in Fig. \ref{fig:x_profile} might imply a possible connection between the thermal and non-thermal processes in the ICM. In Sec. \ref{sec:res_ne}, we find that the discontinuity in the \mbox{X-ray} SB corresponds to a shock Mach number of $\mathcal{M}_{X}=1.52\pm0.05$ if a counter merger shock is present. The true value of the Mach number could be higher since the \mbox{X-ray} emission in the region beyond the SB jump location contains emission from the in-falling materials \citep[e.g.][]{Wang2016}, which might lower the apparent \mbox{X-ray} SB jump and the derived Mach number. In the radio bands, the spatial energy distribution of the radio emission in Fig. \ref{fig:sb_spx_pro} does not indicate a clear spectral steepening in the inner region from the \mbox{X-ray} SB jump as it has been observed in a number of shock-related relics \citep[e.g.][]{Orru2007,Giacintucci2008,Stroe2013a,Bonafede2014,VanWeeren2010a,VanWeeren2016b,VanWeeren2017,Hoang2018a}. This might be due to the large errors associated with the spectral index measurements. However, if the NE radio emission is related to a shock, the spectral index of $-1.08\pm0.12$ at the outermost region would imply a shock Mach number of $\mathcal{M}_\text{NE}=2.1\pm0.2$, which is a higher prediction than the value we obtained from the \mbox{X-ray} data (i.e. $\mathcal{M}_{X}=1.52\pm0.05$).

\section{Conclusions}
\label{sec:conclusions}

We present multi-frequency radio continuum images of the merging galaxy cluster A520 using the new LOFAR 145 MHz data and the archival GMRT 323 MHz and VLA 1.5 GHz data. Combining the multi-frequency radio data sets, we study the morphology and spectral energy distribution of the extended radio emission from the ICM. We also re-analyse the existing \textit{Chandra} \mbox{X-ray} data to search for the possible counter shock in the NE region that was pointed out in \cite{Wang2016}. The main results are described below.

Firstly, we confirm the presence of the large-scale ($760\times950\,\text{kpc}^2$) synchrotron radio emission from the cluster that was detected with the VLA 1.4 GHz observations \citep[e.g.][]{Giovannini1999,Govoni2001c,Vacca2014}. The radio emission in the SW region is enhanced behind the \mbox{X-ray} detected shock. The spectral index measured at the SW radio edge is consistent with the DSA model (acceleration or \mbox{re-acceleration}), assuming the shock Mach number measured with the \mbox{X-ray} observations ($\mathcal{M}_X=2.4_{-0.2}^{+0.4}$; \citealt{Wang2018a}). A spectral steepening is also detected from the SW edge towards the cluster centre. The radio emission is brightest at the steep spectrum region behind the shock front, which is unexpected because of the shock \mbox{(re-)accelerated} electrons age in the downstream region. This might suggest a mixing/projection between halo and the downstream emission from shock \mbox{(re-)accelerated} particles.
        
Secondly, our analysis on the new $20\arcsec$ spectral index map indicates that the extended emission might consist of a radio halo in the centre and, possibly, two other sources in the SW and NE regions. No strong variations of the spectral index of the halo are measured. Furthermore, the radio brightness of the halo is fairly flat and poorly correlated with the \mbox{X-ray} brightness. These facts may suggest that the halo is generated in an extensive turbulent cocoon swept by the outgoing shocks rather than from the very central regions where complex substructures are seen in the \mbox{X-ray} emission.
 
Thirdly, we do not detect extended radio emission in front of the SW shock with the radio observations at 145 MHz, 323 MHz, and 1.5 GHz. This disagrees with the prediction by the gas adiabatic compression model. Our results are in line with a recent study in \cite{Wang2018a}.
        
Finally, we detect an \mbox{X-ray} SB discontinuity in the NE region of the cluster, as was also pointed out in \cite{Wang2016}, which might indicate the presence of a counter merger shock ($\mathcal{M}_\text{NE}^{X}=1.52\pm0.05$). In the same region, a small enhancement of radio emission is visible at high frequencies. If the NE radio emission is shock related, we might detect the steepening of the spectral index behind the \mbox{X-ray} discontinuity. However, we are unable to confirm this owing to the large uncertainty in our spectral measurements.

\section*{Acknowledgments}

We thank the anonymous referee for helpful comments. 
DNH, TS, and HR acknowledge support from the ERC Advanced Investigator programme NewClusters 321271. RJvW acknowledges support from  the VIDI research programme with project number 639.042.729, which is financed by the Netherlands Organisation for Scientific Research (NWO). The LOFAR group in Leiden is supported by the ERC Advanced Investigator programme New-Clusters 321271.  AD acknowledges support by the BMBF Verbundforschung under the grant 05A17STA. FdG is supported by the VENI research programme with project number 639.041.542, which is financed by the Netherlands Organisation for Scientific Research (NWO). 
This paper is based (in part) on data obtained with the International LOFAR
Telescope (ILT) under project code LC7\_025. LOFAR (van Haarlem et al. 2013) is the LOw
Frequency ARray designed and constructed by ASTRON. It has observing, data
processing, and data storage facilities in several countries, which are owned by
various parties (each with their own funding sources), and are collectively
operated by the ILT foundation under a joint scientific policy. The ILT resources
have benefitted from the following recent major funding sources: CNRS-INSU,
Observatoire de Paris, and Université d'Orl\'{e}ans, France; BMBF, MIWF-NRW, MPG,
Germany; Science Foundation Ireland (SFI), Department of Business, Enterprise and
Innovation (DBEI), Ireland; NWO, The Netherlands; The Science and Technology
Facilities Council, UK; Ministry of Science and Higher Education, Poland.
We thank the staff of the GMRT who made these observations possible. The GMRT is run by the National Centre for Radio Astrophysics of the Tata Institute of Fundamental Research.
The National Radio Astronomy Observatory is a facility of the National Science Foundation operated under cooperative agreement by Associated Universities, Inc.
The scientific results reported in this article are based in part on data obtained from the Chandra Data Archive, observations made by the Chandra \mbox{X-ray} Observatory and published previously in cited articles. 
This research has made use of software provided by the Chandra \mbox{X-ray} Center (CXC) in the application packages CIAO, ChIPS, and Sherpa.



\bibliographystyle{aa}
\bibliography{library}

\end{document}